\documentclass[superscriptaddress,amsmath,amssymb,nofootinbib,eqsecnum,a4paper,secnumarabic,preprint,11pt]{revtex4}         
\usepackage[pdftex]{graphicx}
\usepackage{hyperref}     % to make links in equations or refferences.
\usepackage{slashed,color}

\usepackage{titlesec}     % To use \newcommand*{\justifyheading}{\raggedleft}
\newcommand*{\justifyheading}{\raggedright}
\titleformat{\chapter}[display]
  {\normalfont\huge\bfseries\justifyheading}{\chaptertitlename\ \thechapter}
  {20pt}{\Huge}
\titleformat{\section}
  {\normalfont\Large\bfseries\justifyheading}{\thesection}{1em}{}
\titleformat{\subsection}
  {\normalfont\large\bfseries\justifyheading}{\thesubsection}{1em}{}
\titleformat{\subsubsection}
  {\normalfont\bfseries\justifyheading}{\thesubsection}{1em}{}
\usepackage[english]{babel}
\begin{document}

\title{Electric dipole moments and dark matter in a CP violating MSSM}
\author{Tomohiro Abe}
\affiliation{
  Institute for Advanced Research, Nagoya University,
  Furo-cho Chikusa-ku, Nagoya, Aichi, 464-8602 Japan
}
\affiliation{
  Kobayashi-Maskawa Institute for the Origin of Particles and the
  Universe, Nagoya University,
  Furo-cho Chikusa-ku, Nagoya, Aichi, 464-8602 Japan
}
\author{Naoya Omoto}
\affiliation{
Department of Physics, Hokkaido University, Sapporo 060-0810, Japan
}
\author{Osamu Seto}
\affiliation{
Institute for International Collaboration, Hokkaido University, Sapporo 060-0815, Japan
}
\affiliation{
Department of Physics, Hokkaido University, Sapporo 060-0810, Japan
}
\author{Tetsuo Shindou}
\affiliation{
Division of Liberal-Arts, Kogakuin University, 
Nakano-machi, Hachioji, Tokyo, 192-0045 Japan
}

\begin{abstract}
We investigate electric dipole moments (EDMs) in a CP-violating minimal supersymmetric standard model  with the Bino-like neutralino dark matter (DM) annihilating through the heavy Higgs funnel. Motivated by the current experimental results, in particular, the measured mass of the standard model-like Higgs boson, we consider a mass spectrum with stop masses of about 10 TeV. For the other sfermions, we consider masses of about 100 TeV. We show that CP-violating phases of the order of ten degrees in gaugino and Higgsino mass parameters are consistent with the current bound by EDMs of the electron, the neutron, and the mercury. They are within the reach of future experiments. We also show that effects of CP-violating phases induce a difference in DM-nucleon scattering cross section by a factor.
\end{abstract}

\preprint{EPHOU-18-005}
\preprint{KU-PH-022} 

%\date{September 1, 2017 \\ \ \  Revised: \today}

\maketitle
%\tableofcontents
% End title 
%======================================================================

\section{Introduction}

Supersymmetry is an attractive candidate of physics beyond the Standard Model (SM), although current results from LHC experiments indicate that supersymmetric (SUSY) particles are heavier than have been expected. Attractive aspects come from the fact that, for instance, the gauge coupling unification is realized in the minimal SUSY SM (MSSM), the gauge hierarchy problem is improved, and elementary scalar fields such as Higgs fields are introduced in a theoretically natural way. Moreover, SUSY models may provide additional interesting consequences. SUSY interpretation of muon anomalous magnetic moment is one example~\cite{Lopez:1993vi,Chattopadhyay:1995ae,Moroi:1995yh}. SUSY models contain new sources of CP violation and/or flavor violation, which potentially induce new CP or flavor violating phenomena. The lightest SUSY particle (LSP) is stable and hence a good candidate for dark matter (DM) in our Universe, if the R-parity is unbroken~\cite{Goldberg:1983nd,Ellis:1983ew}. 

Electric dipole moments (EDMs) of the neutron and other heavy atoms
 are prime physical quantities for probing sources of CP violation.
Parameters in the MSSM generally pose several CP violating phases. 
It used to be regarded that the null experimental EDM results confront 
 the MSSM with $\mathcal{O}(1)$ CP violating phases and $\mathcal{O}(100)$~GeV masses of SUSY
 particles~\cite{Ellis:1982tk,Buchmuller:1982ye,Polchinski:1983zd,Dugan:1984qf}. 
The LHC results suggest that masses of many SUSY particles are larger than $\mathcal{O}(10)$~TeV.
\footnote{
To be precise, a scenario with SUSY particles with masses of a few TeV is still allowed.
The current limit on gluino mass is around 2~TeV and the squark masses can be 
smaller than 3~TeV in the degenerate case.
}
Therefore CP violating phases of order unity in the SUSY sector~\cite{Nath:1991dn,Kizukuri:1991mb} seems still likely and worth investigating.
For recent studies, see e.g. Refs~\cite{Arbey:2016cey,Berger:2015eba}.

In the MSSM with R-parity, the lightest neutralino $\tilde{\chi}$ is a candidate of 
 the weakly interacting mass particle (WIMP) DM. 
While LHC experiments as well as direct detection experiments of DM, such as
 LUX~\cite{Akerib:2013tjd,Akerib:2016vxi}, XENON1T~\cite{1705.06655}, and PandaX-II~\cite{Tan:2016zwf,1708.06917}
 are constraining large parameter space of the MSSM, 
 there are still viable scenarios reproducing thermal relic abundance of the DM consistently.
Appropriate magnitude of annihilation cross section of neutralino in the early Universe is realized 
 if (i) neutralinos annihilate significantly through $\text{SU}(2)$ gauge interaction, 
 or (ii) annihilation cross section of Bino-like neutralino is enhanced
 with a particular mass spectrum of other associated particles.

Higgsino-like neutralino DM with the mass of about $1$~TeV is an example in the former class.
Phenomenology in this scenario such as the direct detection of DM,
 contribution to the EDMs, and collider signals have been precisely
 studied in Ref.~\cite{Nagata:2014wma}.

In this paper, we focus on another case in the later class;
 a Bino-like neutralino DM annihilates through heavy Higgs boson resonance~\cite{Drees:1992am,Baer:1995nc,Baer:1997ai,Barger:1997kb,Ellis:2001msa}.\footnote{For a study of CP violation in stau coannihilation scenario, see e.g., Ref.~\cite{Belanger:2008yc}} 
In this scenario where heavy Higgs boson resonance in the neutralino DM annihilation is utilized,
 masses of the heavy Higgs boson are about twice of the mass of the neutralino DM.
Since the Bino-like neutralino contains small Higgsino component,
 the neutralino can be searched through Higgs bosons exchange by spin-independent scattering off nucleus~\cite{Jungman:1995df}.
Masses of stops would be around $10$~TeV
 in order to reproduce the measured SM-like Higgs boson mass ($m_h=125$~GeV)~\cite{Draper:2016pys,Allanach:2018fif}.
Then, all the other SUSY particle masses and parameters except for the Bino mass,
 the Higgsino mass parameter $\mu$, $B\mu$, and stop masses
 can be much larger than $\mathcal{O}(1)$~TeV.
With such SUSY particle mass spectrum,
 most of SUSY contributions to the low energy phenomena can be decoupled
 as the irrelevant SUSY particles are heavier, 
 SUSY contributions to the EDMs in CP violating models
 can still be significantly large nevertheless.
The main goal of this article is, by decoupling the other particles,
 to estimate the magnitude of EDMs induced by CP violation in neutralino DM sector
 with taking account of CP-violating phase effects into the thermal DM
 abundance~\cite{Falk:1995fk,Gondolo:1999gu,Gomez:2004eka,Nihei:2004bc,Choi:2006hi,Belanger:2006qa}.

We examine the electron EDM, the nucleon EDM, and the mercury EDM,
 as well as the DM-nucleon scattering cross section on the parameter space,
 where appropriate thermal DM relic abundance is reproduced, 
 for order unity CP-violating phases of gaugino mass parameters and the $\mu$ parameter.
For non-vanishing CP phase of $\mu$ and $A$ parameters, 
 see, for example, Refs.~\cite{Falk:1995fk,Nihei:2004bc,hep-ph/0311314,hep-ph/0506106}.
The magnitude of scattering off cross section between DM and nucleon
 is affected by the CP violating
 phases~\cite{Chattopadhyay:1998wb,Falk:1998xj,Falk:1999mq,Choi:2000kh,Nihei:2004bc, Abe:2014gua, Abe:2017glm}. 
We also study the dependence of spin-independent cross section of the DM in our scenario
 and find that the effect changes by a factor. 
We show that wide parameter regions in our scenario are now unconstrained yet,
 but will be explored by future experiments. 

This article is organized as follows.
In Sec.~\ref{SecSetup}, we define a benchmark scenario for studying phenomenology in our DM scenario.
In Sec.~\ref{SecObs}, we show the results of our analysis on several EDM measurements and the spin-independent cross section. Summary and conclusion are presented in Sec.~\ref{SecSummary}.

\section{Setup of the scenario}
\label{SecSetup}

In this section, we briefly review the MSSM Lagrangian, and we describe the parameter setup for our analysis. 
The superpotential and the soft SUSY breaking terms in the MSSM are given by~Ref.~\cite{Allanach:2008qq} 
\begin{equation}
W=\epsilon_{ab}\left[(y_e)_{ij}H_1^aL_i^b\bar{E}_j
+ (y_d)_{ij}H_1^aQ_i^b\bar{D}_j
+ (y_u)_{ij}H_2^aQ_i^b\bar{U}_j
-\mu H_1^aH_2^b\right]\;,
\end{equation}
and 
\begin{align}
\mathcal{L}_{\text{soft}}
=&\ -\frac{M_1}{2}\tilde{B}\tilde{B}-\frac{M_2}{2}\tilde{W}^{\alpha}\tilde{W}^{\alpha}-\frac{M_3}{2}\tilde{G}^A\tilde{G}^A\nonumber\\
&\ 
-m_{H_1}^2H_{1a}^*H_1^a+m_{H_2}^2H_{2a}^*H_2^a
-\tilde{q}_{iLa}^*(M_{\tilde{q}}^2)_{ij}\tilde{q}_{jL}^a
-\tilde{\ell}_{i_La}^*(M_{\tilde{\ell}}^2)_{ij}\tilde{\ell}_{jL}^a\nonumber\\
&\ -\tilde{u}_{iR}(M_{\tilde{u}}^2)_{ij}\tilde{u}_{jR}^{*}
-\tilde{d}_{iR}(M_{\tilde{d}}^2)_{ij}\tilde{d}_{jR}^{*}
-\tilde{e}_{iR}(M_{\tilde{e}}^2)_{ij}\tilde{e}_{jR}^{*}\nonumber\\
&\ 
-\epsilon_{ab}\left[(T_e)_{ij}H_1^a\tilde{\ell}_{iL}^b\tilde{e}_{jR}
+(T_d)_{ij}H_1^a\tilde{q}_{iL}^b\tilde{d}_{jR}
+(T_u)_{ij}H_2^a\tilde{q}_{iL}^b\tilde{u}_{jR}
+m_3^2H_1^aH_2^b+\text{h.c.}\right]\;,
\end{align}
respectively.
The convention of the epsilon tensor is $\epsilon_{12}=-\epsilon_{21}=1$.
Here, we note that gaugino mass parameter for Bino $M_1$, Wino $ M_2$, and gluino $M_3$ are in general complex. 
In the following, we focus on the Yukawa couplings of the third generation quarks and leptons, so that 
we use $y_t$, $y_b$, and $y_{\tau}$ for the Yukawa couplings of top, bottom, and tau, respectively.  
Neglecting the flavor mixing in the soft SUSY breaking terms, we take flavor diagonal soft scalar masses
 as $M_{\tilde{q}_i}^2=(M_{\tilde{q}}^2)_{ii}$, $M_{\tilde{\ell}_i}^2=(M_{\tilde{\ell}}^2)_{ii}$, 
$M_{\tilde{u}_i}^2=(M_{\tilde{u}}^2)_{ii}$, $M_{\tilde{d}_i}^2=(M_{\tilde{d}}^2)_{ii}$, and 
$M_{\tilde{e}_i}^2=(M_{\tilde{e}}^2)_{ii}$. For the trilinear couplings, $A$ parameters defined by 
$(T_u)_{33}=A_{\tau}y_{t}$, 
$(T_d)_{33}=A_{\tau}y_{b}$, and $(T_e)_{33}=A_{\tau}y_{\tau}$ are used.

In the MSSM, the mass of the SM-like Higgs boson is expressed with some SUSY breaking parameters.
In our analysis, we take $\tan\beta:= \langle H_2\rangle/\langle H_1\rangle =30$ and 
we fix the stop mass parameters as 
$M_{\tilde{q}_3} = 7~\text{TeV}$, $M_{\tilde{t}}:= M_{\tilde{u}_3} = 7~\text{TeV}$ and $A_{t} = 10~\text{TeV}$,
then the measured SM-like Higgs boson mass $m_h\simeq 125$~GeV can easily reproduced~\cite{Draper:2016pys,Allanach:2018fif}.
The other SUSY particles are relevant to neither the mass of the SM-like Higgs boson nor the DM relic density.
We may assume that those are much heavier than stop so that
 those are decoupled from low energy observables. 
We here take masses of the other sfermions as $100$~TeV and the Wino and gluino masses to be $10$~TeV. 
In this article, we focus on the Bino-like DM with the Higgs funnel scenario,
 where heavy Higgs boson mass is close to twice the mass of the DM so that 
 the Bino-like neutralino rapidly annihilate through the heavy Higgs bosons resonance and
 has left with the appropriate cosmic abundance for DM. 
Since masses of heavier neutral Higgs bosons, $m_H$ and $m_A$, are close to the charged Higgs boson mass $m_{H^{\pm}}$ in the MSSM,
 we fix $m_{H^{\pm}}$ to be twice of Bino mass parameter $M_1$ to realize resonant annihilation
 by the heavy Higgs bosons.
In addition, the $\tilde{\chi}$-$\tilde{\chi}$-Higgs boson coupling depends
 on non-vanishing Higgsino component in the neutralino.
Thus, both the Bino mass $|M_1|$ and the Higgsino mass $|\mu|$ should be of the order of TeV. 
We leave $M_1$ as a free parameter and solve $|\mu|$ from the measured dark matter energy density.
We summarize the parameter set in our analysis as follows:
\begin{align}
& |M_2| = |M_3| = 10~\text{TeV}, 
\label{eq:Bench_1}
\\
& M_{\tilde{q}_{1,2}} = M_{\tilde{u}_{1,2}} = M_{\tilde{d}_{1,2,3}} 
=M_{\tilde{\ell}_{1,2,3}}=M_{\tilde{e}_{1,2,3}}= 100~\text{TeV},
\label{eq:Bench_2}
\\
& M_{\tilde{q}_3} = M_{\tilde{t}} = 7~\text{TeV},
\label{eq:Bench_3}
\\
& A_{t} = 10~\text{TeV},
\label{eq:Bench_4}
\\
& m_{H^{\pm}} = 2 M_1,
\label{eq:Bench_5}
\\
& \tan\beta = 30.
\label{eq:Bench_6}
\end{align}
The other $A$-terms are zero.
With the above parameter set,
 besides the CKM phase and the CP phases in the sfermion mass matrices,
 five parameters, $\mu$, gaugino masses $M_i$ and $A_t$, may have CP phases
 $(\phi_{\mu}, \ \phi_{M_1}, \ \phi_{M_2}, \ \phi_{M_3}, \ \phi_{A_t})$, respectively.
Here, each phases of a quantity $X$ are defined by $ X = |X| e^{i \phi_X}$.

There is a rephasing degree of freedom in the MSSM. 
Thus, all the physical quantities are described by the following combinations,
\begin{equation}
\arg(M_iM_j^*)\;,\quad 
\arg(M_iA_t^*)\;,\quad 
\arg(\mu M_i)\;,\quad
\arg(\mu A_t)\;,\quad (i,j=1,2,3)\;.	
\end{equation}
Without loss of generality, we can take the basis of CP phases as $\phi_{M_3}=0$.
In addition, we take $\phi_{A_t} =0$ to concentrate on the CP violation in the neutralino sector as well as, technically speaking, to keep $m_h \simeq 125$ GeV avoiding complicated parameter dependence of the SM-like Higgs boson mass.
In general, the non-zero value of $\phi_{A_t}$ significantly contributes to the predictions of the EDMs. 
However, in our parameter set given in Eqs.~(\ref{eq:Bench_1}
) -- (\ref{eq:Bench_6}), we find the contribution from $\phi_{A_t}$ is negligible because the mass splitting between two stops is small.
Therefore, we here set $\phi_{A_t}=0$, and 
we scan the following four parameters,
%The parameters associated with coloured particles, namely $\phi_{A_t}$ affect
%to the Higgs potential. 
%
%
%
\begin{align}
 (|M_1|,\ \phi_{\mu}, \ \phi_{M_1},\  \phi_{M_2}).
\end{align}

\section{Observables}
\label{SecObs}

As we mentioned in the previous section, we choose $|\mu|$ to achieve the correct DM relic density
as $\Omega_{\text{DM}}h^2=0.1198 \pm 0.0015$~\cite{1502.01589}.
We use \texttt{micrOMEGAs 4.3.5}~\cite{1606.03834} with \texttt{CPsuperH2.3}~\cite{arXiv:1208.2212} 
in calculations of dark matter thermal relic density and the Higgs mass.
In our benchmark point, the Higgs mass is almost fixed to be 125~GeV. There is small fluctuation 
of order of 0.1~GeV by scattering the parameters. On the other hand, the calculation of the Higgs mass has 
theoretical uncertainty of order of a few GeV. So we consider that our benchmark points are 
consistent with measurements of the Higgs mass at the LHC. 

With the correct DM relic abundance and the correct Higgs mass, 
we calculate the electron EDM, the neutron EDM, and the mercury EDM.
The electron and mercury EDMs give strong constraints on the parameter space as we will see later.
We also discuss the scattering cross section for the direct detection experiments.

\subsection{EDM}
The EDMs of fermions ($d_f$), the EDM of electron ($d_e$), the chromo EDM (cEDM) of quarks ($d^C_q$),
 and the Wilson coefficient of the Weinberg operator ($\omega$) are defined by 
\begin{align}
 {\cal L}
\supset&
 - d_f \frac{i}{2} \bar{f} \sigma^{\mu \nu} \gamma_5 f F_{\mu \nu}
 %- g_s d^C_q \frac{i}{2} \bar{q} \sigma^{\mu \nu} \gamma_5 q F_{\mu \nu}
 - g_s d^C_q \frac{i}{2} \bar{q} \sigma^{\mu \nu} \gamma_5 q G_{\mu \nu}
 - \omega \frac{1}{6} f^{abc} G^{a}_{\mu \nu} G^{b}{}^{\nu}_{\ \rho} G^{c}_{\alpha \beta} \epsilon^{\rho \mu \alpha \beta},
\end{align}
where the convention of the epsilon tensor is $\epsilon^{0123}=+1$.
We calculate $d_u$, $d_d$, $d_e$, $d^C_u$, and $d^C_d$
by using \texttt{CPsuperH2.3}~\cite{arXiv:1208.2212} implemented in \texttt{micrOMEGAs 4.3.5}~\cite{1606.03834}. 
We use the formulae given in Ref.~\cite{1712.09503} and couplings calculated by \texttt{CPsuperH2.3} 
to evaluate $\omega$.\footnote{\texttt{CPsuperH2.3} also calculate the Wilson coefficient of the Weinberg operator, 
but it returns very unstable numbers during the scanning the parameter space because of the loss of significant digits.} 
These EDMs and the Wilson coefficient are evaluated at the electroweak scale $\mu_W = m_t$.
The neutron EDM and the mercury EDM have to be evaluated at the hadronic scale ($\mu_H \simeq 1$~GeV). 
The renormalization group evolution from the electroweak scale to the hadronic scale is taken into account~\cite{hep-ph/0510137}.
At the leading order of QCD, we find\footnote{The definition of the cEDM
 in the \texttt{CPsuperH2.3} is different from ours, $\left. d^C_q  \right|_{\texttt{CPsueprH2.3}} = g_s d^C_q.$}
\begin{align}
\frac{d_u}{e}(\mu_H)   =& 0.35 \frac{d_u}{e}(\mu_W) -0.17 g_s(\mu_W) d^C_u(\mu_W) - (9.24874 \times 10^{-5} \text{GeV}) \omega(\mu_W), \\
d^C_u(\mu_H) =& 0.34 g_s(\mu_W) d^C_u(\mu_W) + (0.00031 \text{GeV})\omega(\mu_W), \\
\frac{d^e_d}{e}(\mu_H) =& 0.40 \frac{d_u}{e}(\mu_W) +0.098  g_s(\mu_W)  d^C_d(\mu_W) +(0.00010 \text{GeV}) \omega(\mu_W),\\
d^C_d(\mu_H) =& 0.38 g_s(\mu_W) d^C_d(\mu_W) + (0.00070 \text{GeV})\omega(\mu_W),\\
\omega(\mu_H) =&  0.39 \omega(\mu_W).
\end{align} 
Here the unit of the EDMs and of the cEDMs are GeV$^{-1}$, and the unit of $\omega$ is GeV$^{-2}$.
In the evaluation, we used the following values,
\begin{align}
& g_s(\mu_W) = 1.1666,\ g_s(\mu_H)=2.13309,\\
%& e(\mu_W) = 0.302822,\ e(\mu_H)=0.313535,\\
& m_u(\mu_W)=0.003~\text{GeV},\ m_u(\mu_H)=0.0024~\text{GeV},\\ 
& m_d(\mu_W)=0.006~\text{GeV},\ m_d(\mu_H)=0.0054\text{GeV}. 
\end{align}
We estimate the neutron EDM and the Mercury EDM as 
\footnote{
Some theoretical uncertainty in the EDM calculations are known. 
The neutron EDM calculation has uncertainty of factor two~\cite{Hisano:2014mna}, 
and the mercury EDM calculation has $20-30 \%$ uncertainty~\cite{Ginges:2003qt}.
}
\begin{align}
d_n=& 0.79 d_d-0.20 d_u+ e (0.30 d^C_u+0.59 d^C_d) \pm (10-30)\text{MeV} \omega,
\label{eq:dn}
\\
%\frac{d_{Hg}}{e}=& -1.8 \times 10^{-4} \times (4^{+8}_{-2}) \times (d^C_u-d^C_d).
\frac{d_{Hg}}{e}=& 7 \times 10^{-3} \times (d^C_u-d^C_d) - 10^{-2} \frac{d_e}{e} .
\end{align}
Here, we used the results given in Refs.~\cite{hep-ph/0208257,1308.6493} for $d_n$
and the results given in Refs.~\cite{hep-ph/0311314,hep-ph/0506106,0808.1819} for $d_{Hg}$.
We found the contribution from the Weinberg operator is much smaller than the contributions from cEDMs. 
Thus, we ignore the contribution from the Weinberg operator in the following analysis.

Upper bounds on the electron EDM~\cite{1310.7534},
the mercury EDM~\cite{1601.04339},
and the neutron EDM~\cite{hep-ex/0602020} are
\begin{align}
 |d_{e}| <& 8.7 \times 10^{-29}~\text{e}\cdot \text{cm}  \quad \text{(90\% C.L.)},\\
 |d_{Hg}| <& 7.4 \times 10^{-30}~\text{e}\cdot \text{cm}  \quad \text{(95\% C.L.)},\\
 |d_{n}| <& 2.9 \times 10^{-26}~\text{e}\cdot \text{cm}  \quad \text{(90\% C.L.)}.
\end{align}
We note that the constraint from the thallium EDM~\cite{Regan:2002ta} is in practice equivalent
 with that to the electron EDM~\cite{LiuKelly,Martensson-Pendrill,Lindroth,Pospelov:2005pr}.
Prospects for the electron and neutron EDMs are
\begin{align}
|d_e| =& 1 \times 10^{-30}~\text{e}\cdot\text{cm~~\cite{arXiv:1208.4507,Kawall:2011zz}}, \label{eq:prospect_de}\\
|d_n| =& 2.5 \times 10^{-29}~\text{e}\cdot\text{cm~\cite{1201.5773}}.  \label{eq:prospect_dn} 
\end{align}
% $|d_e| = 1 \times 10^{-30)} $e cm \cite{arXiv:1208.4507,Kawall:2011zz}
% $|d_n| = 2.5 \times 10^{-29)} $e cm \cite{1201.5773}
%
% $|d_e| = 1 \times 10^{-29)} $e cm \cite{Sakemi:2011zz}
%  $|d_n| = 1.7 \times 10^{-28)} $e cm \cite{0709.2428}
%  $|d_n| = 5 \times 10^{-28)} $e cm [26]
% [26] I. Altarev et al., Nucl. Instrum. Meth. A 611, 133 (2009). doi:10.1016/j.nima.2009.07.046
%
%  $|d_n| = 2.5 \times 10^{-29)} $e cm \cite{1201.5773}
% [27] A. Lehrach, B. Lorentz, W. Morse, N. Nikolaev and F. Rathmann, arXiv:1201.5773 [hep-ex].
%
%
%  $|d_n| = 5 \times 10^{-28)} $e cm [26]
% [26] K. Kirch, http://vmsstreamer1.fnal.gov/Lectures/Colloquium/presentations/130213Kirch.pdf
%%%
%%% edited by Shindou

\begin{figure}[ht]
\begin{center}
\includegraphics{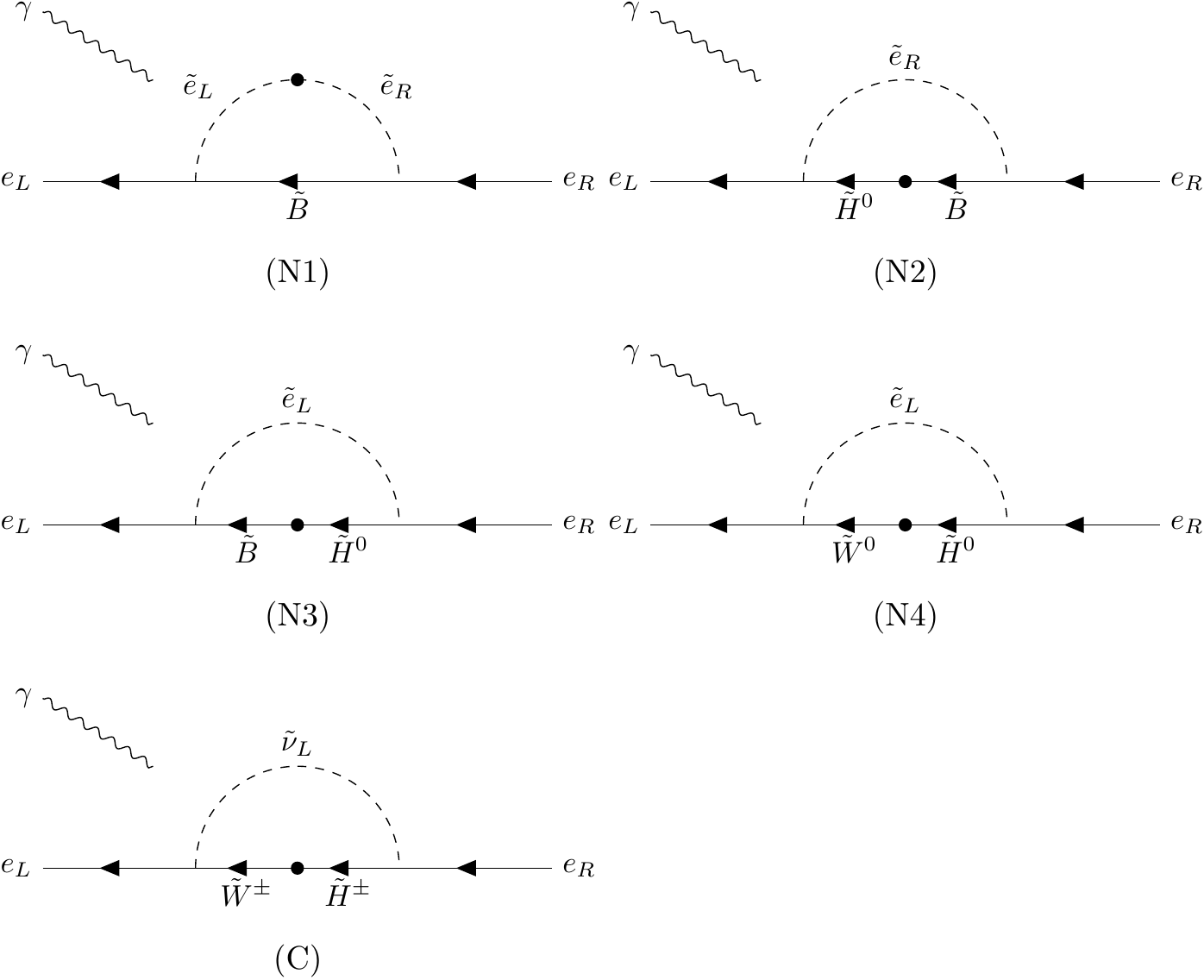}
\end{center}
\caption{Diagrams for leading-order contributions 
to the electron EDM at the one-loop level. \label{fig:MIemdOneloop}}
\end{figure}

\begin{figure}[ht]
\begin{center}
\includegraphics{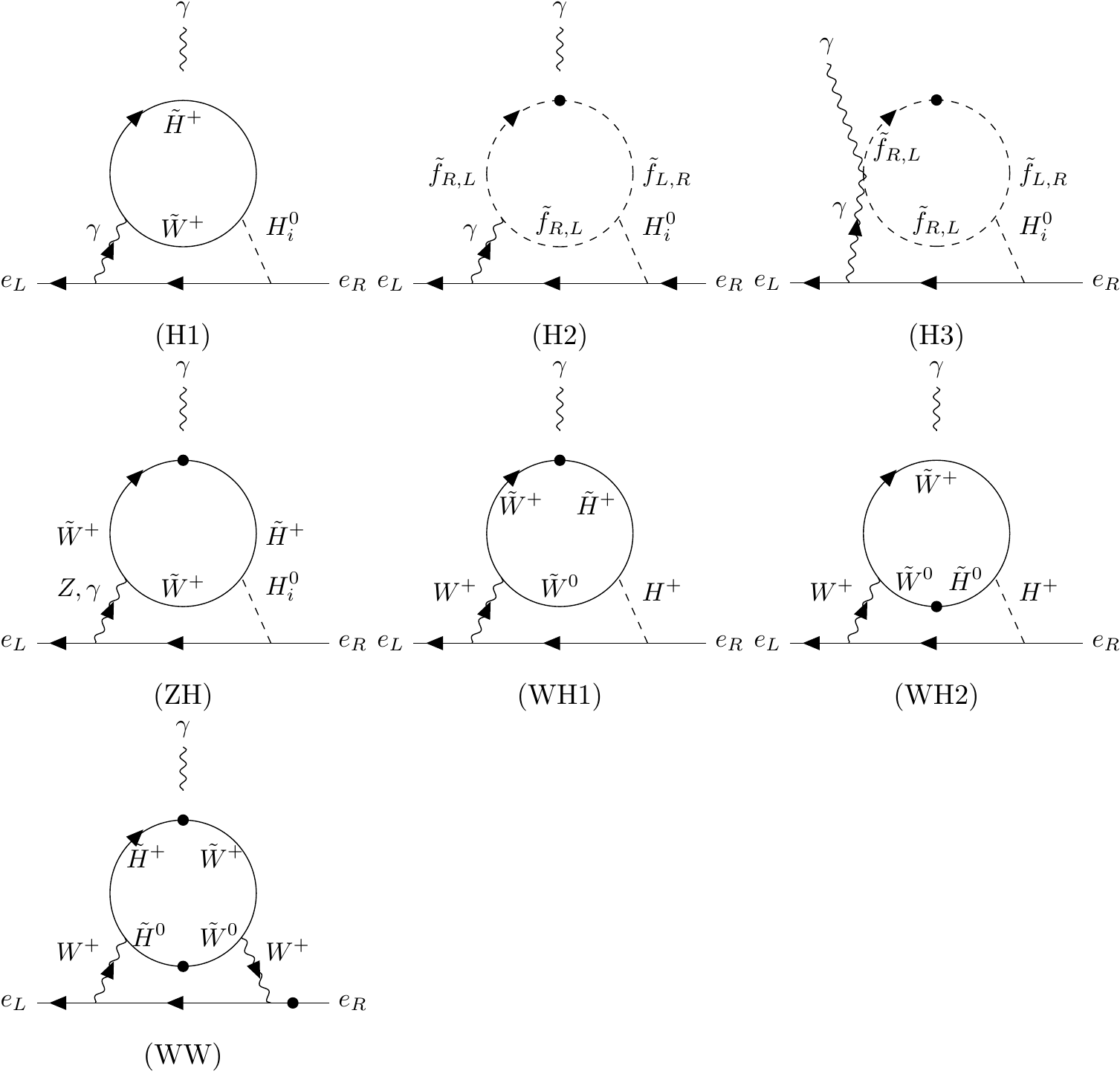}
\end{center}
\caption{Diagrams for leading-order contributions 
to the electron EDM at the two-loop level. 
In (H2) and (H3), stop and sbottom contributions are larger than 
the other sfermion contributions.  
\label{fig:MIBarrZee}}
\end{figure}

Let us consider the parameter dependence of the electron EDM in the MSSM. 
The quark EDM has the similar dependence to the electron EDM.
% The chromo EDM has different dependence due to the running effect.
In our setup, the EDMs are generated by the diagrams shown in Figs.~\ref{fig:MIemdOneloop} and \ref{fig:MIBarrZee}.
First, we see the one-loop contributions, where SUSY contributions are 
given by slepton-electroweakino loop diagrams.
The relevant diagrams which give leading order contributions are shown in  
Fig.~\ref{fig:MIemdOneloop}.
The contribution from each diagram depends on the SUSY parameter as follows:
\begin{align}
&d_e^{(\text{N1})}\propto \mathrm{Im}(M_{eLR}^2 M_1)
\;,\quad 
d_e^{(\text{N2})}\propto \mathrm{Im}(\mu M_1)\;,\quad 
d_e^{(\text{N3})}\propto \mathrm{Im}(\mu M_1)\;,\quad 
d_e^{(\text{N4})}\propto \mathrm{Im}(\mu M_2)\;,\quad \nonumber\\
&d_e^{(\text{C})}\propto \mathrm{Im}(\mu M_2)\;,
\end{align}
 with
\begin{equation}
M_{eLR}^2=A_e^{*}m_e-\mu m_e \tan\beta\;.
\end{equation}
%
%
%
%\section{Barr-Zee diagrams}
%
At the two-loop level, Barr-Zee diagrams shown in Fig.~\ref{fig:MIBarrZee}
give significant contributions.
We can see that (WW) diagram is strongly suppressed compared to the other diagrams because it has not only a chirality suppression 
by the electron mass but also a suppression by double mass insertion in the loop.
Contribution from each diagram depends on the SUSY parameter as follows:
\begin{align}
&d_e^{(\text{H1})}\propto \mathrm{Im}(\mu M_2)\;,\quad 
d_e^{(\text{H2})}\propto \mathrm{Im}(M_{fLR}^{2*}\mu)\;,\quad
d_e^{(\text{H3})}\propto \mathrm{Im}(M_{fLR}^{2*}\mu)\;,\quad\nonumber\\
&d_e^{(\text{ZH})}\propto \mathrm{Im}(\mu M_2)\;,\quad 
d_e^{(\text{WH1})}\propto \mathrm{Im}(\mu M_2)\;,\quad 
d_e^{(\text{WH2})}\propto \mathrm{Im}(\mu M_2)\;,\quad 
d_e^{(\text{WW})}\propto \mathrm{Im}(\mu M_2)\;,
\end{align}
 with 
\begin{equation}
M_{fLR}^2=\begin{cases}
A_t^*m_t-m_t\mu \cot\beta&f=t\;,\\
A_f^*m_f-m_f\mu \tan\beta&f=b,\tau\;.
\end{cases}
\end{equation}
%Note that the EDM (electron, quark, chromo) depends on the phases of 
%the following combinations of the MSSM parameters:
%\begin{equation}
%\arg(M_iM_j^*)\;,\quad 
%\arg(M_iT_f^*)\;,\quad 
%\arg(\mu M_i)\;,\quad 
%\arg(\mu T_f)\;.
%\end{equation}


In our benchmark case, the one-loop contributions are strongly suppressed by large selectron and sneutrino masses, and 
the two-loop Barr-Zee diagrams provide dominant 
contributions unless Wino, stop, and sbottom masses are heavy enough to be decoupled. 
In the diagrams (H2) and (H3), the stop and sbottom loops dominate the contribution, because the color factor enhances it and the Yukawa couplings are larger than the other sfermions.
Therefore, in general, the CP phase of $A_t$ gives a significant contribution to the EDMs through these two diagrams. 
However the contribution of $\phi_{A_t}$ is highly suppressed in our benchmark case, because
 the contributions of the stop loop to the EDMs is proportional to the
$\tilde{t}_L$-$\tilde{t}_R$ mixing, $\theta_{\tilde{t}}$, and thus
\begin{align}
 \left. d \right|_{\text{stop}} \propto
 \sin\theta_{\tilde{t}} \cos\theta_{\tilde{t}} \left( f(m_{\tilde{t}_1}) - f(m_{\tilde{t}_2})  \right),
\end{align}
 where $f$ is a loop function depending on the stop mass in the loop.
Thanks to the relative sign of the mixing angle, the contributions from $\tilde{t}_1$ and $\tilde{t}_2$ 
are destructive. As a result, the EDMs are insensitive to $\phi_{A_t}$ for the degenerated
stops masses regime.  
For example, the contribution to electric dipole moment $d_e/e$ is of the order of $10^{-30}$ cm for $\phi_{A_t}=30^{\circ}$, $\phi_{\mu}=\phi_{M_1}=\phi_{M_2}=0$, and $|M_1|=1~\text{TeV}$. 
It is one order of magnitude below the present experimental bound, but it is within the reach of future experiment. 
In the diagrams (H1), (ZH), (WH1) and (WH2), the leading 
order contributions are given by the Wino and the Higgsino loops
so that these diagrams are decoupled when 
the Wino mass $M_2$ becomes larger. 
%
% End: edited by Shindou
%

\begin{figure}[tb]
\includegraphics[width=0.38\hsize]{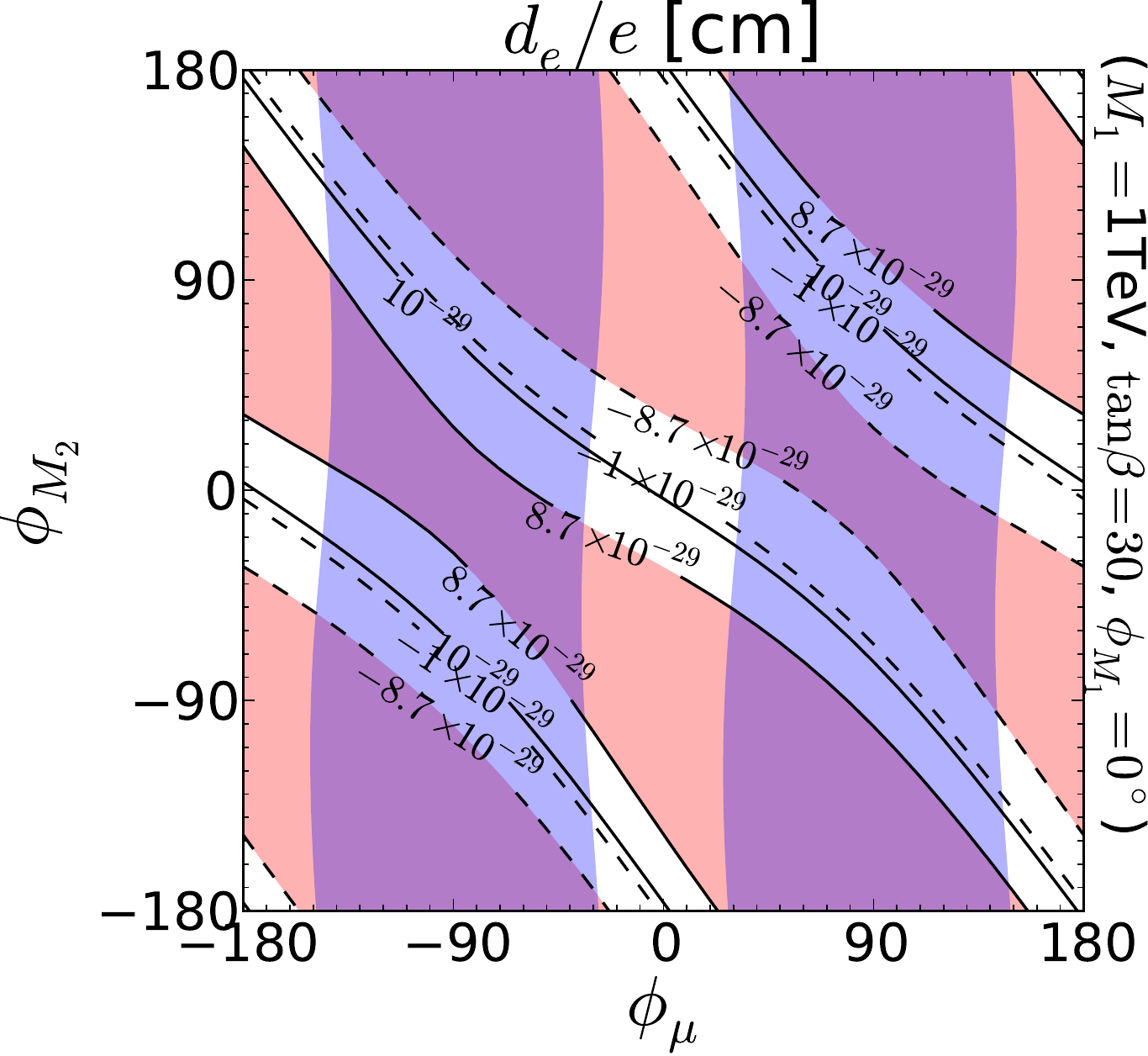}
\quad
\includegraphics[width=0.38\hsize]{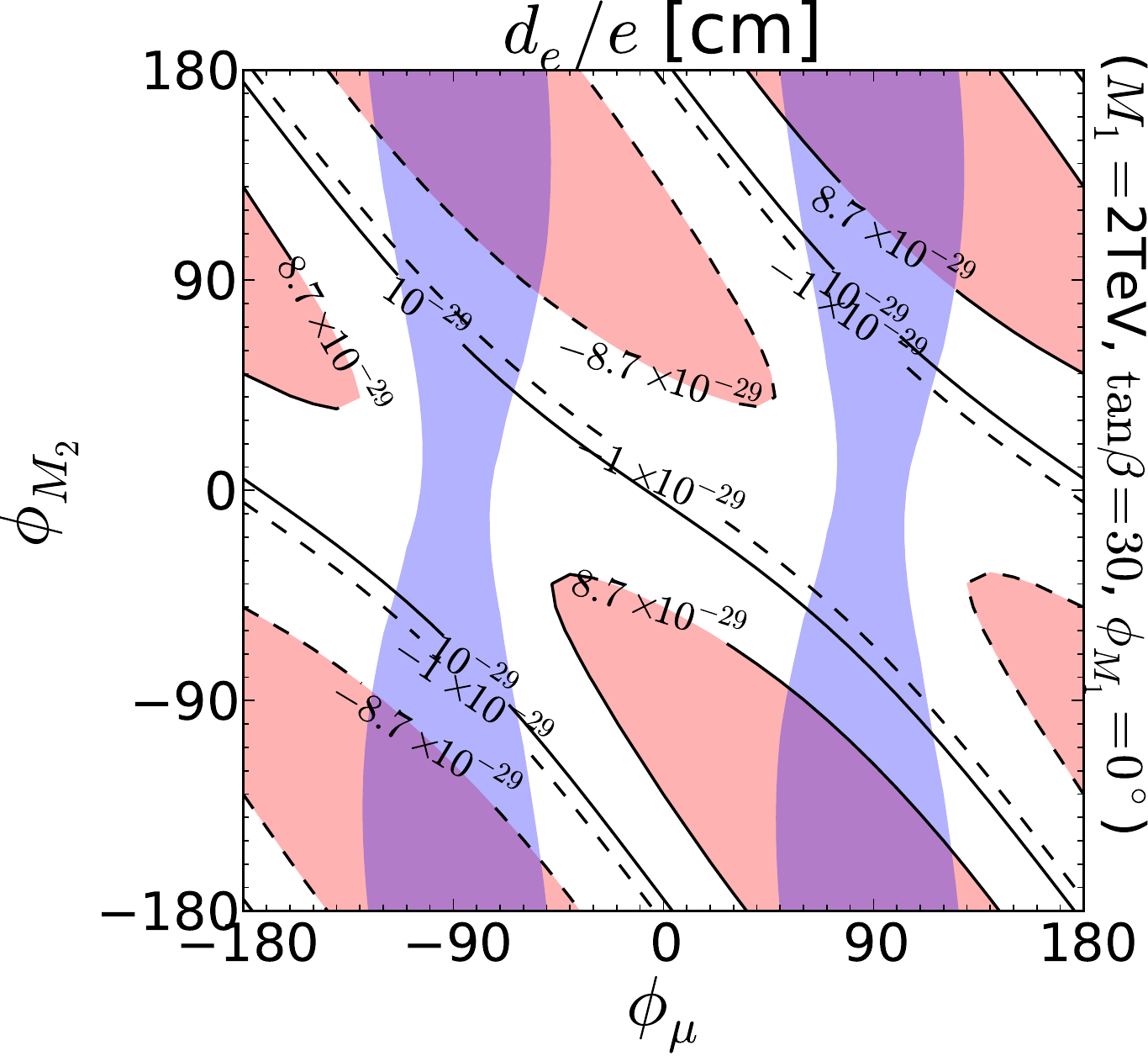}
\\
\includegraphics[width=0.38\hsize]{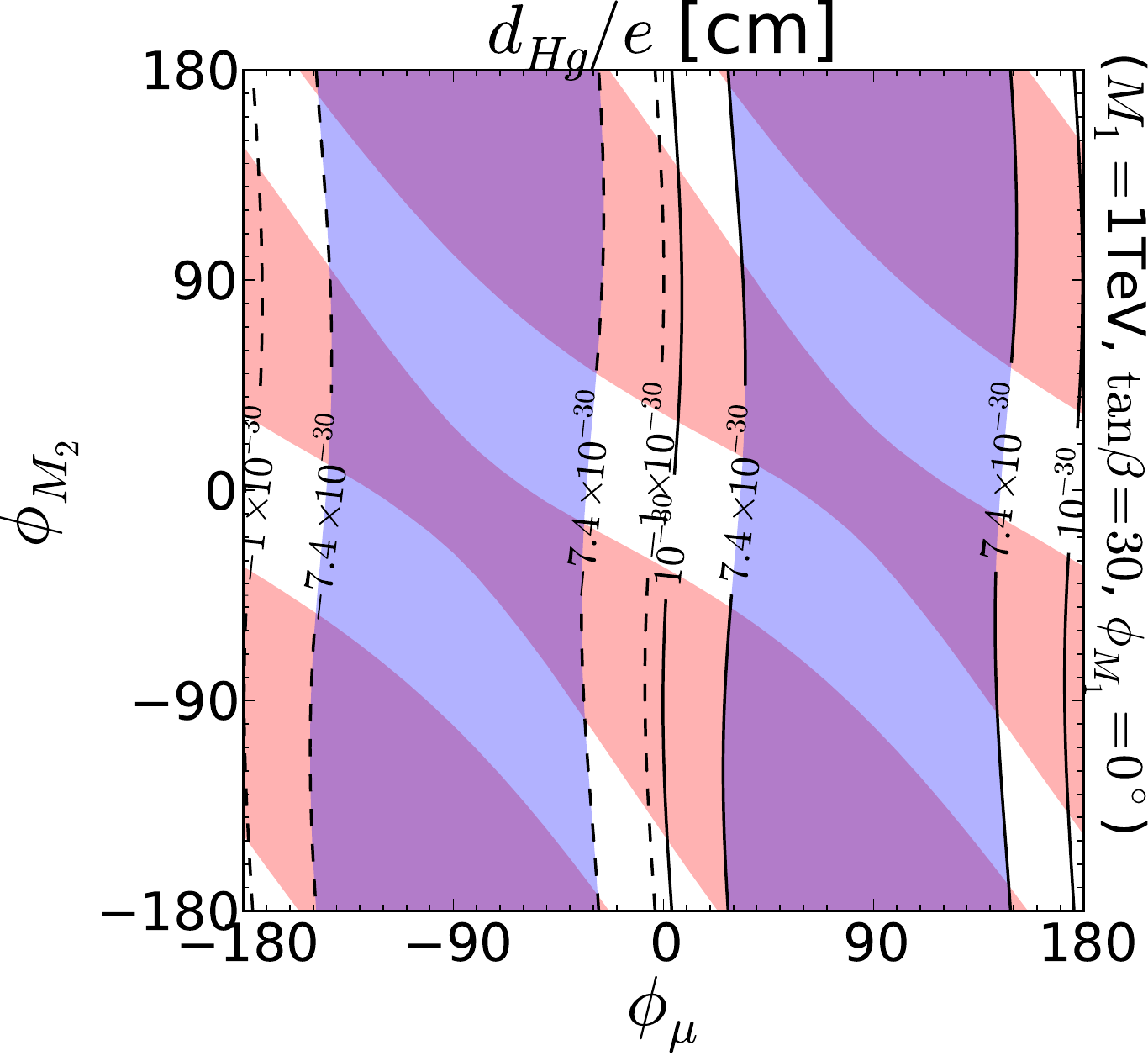}
\quad
\includegraphics[width=0.38\hsize]{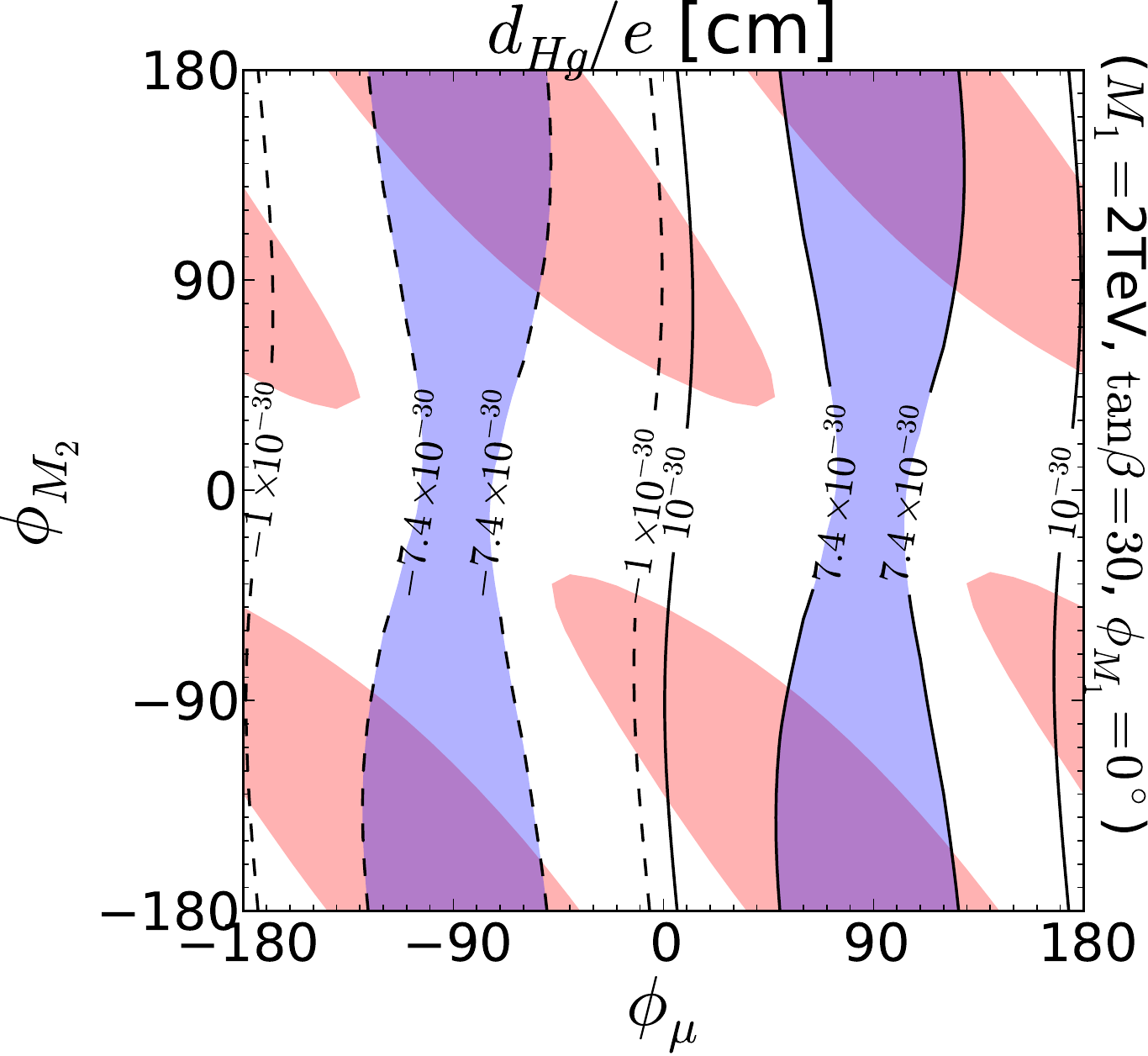}
\\
\includegraphics[width=0.38\hsize]{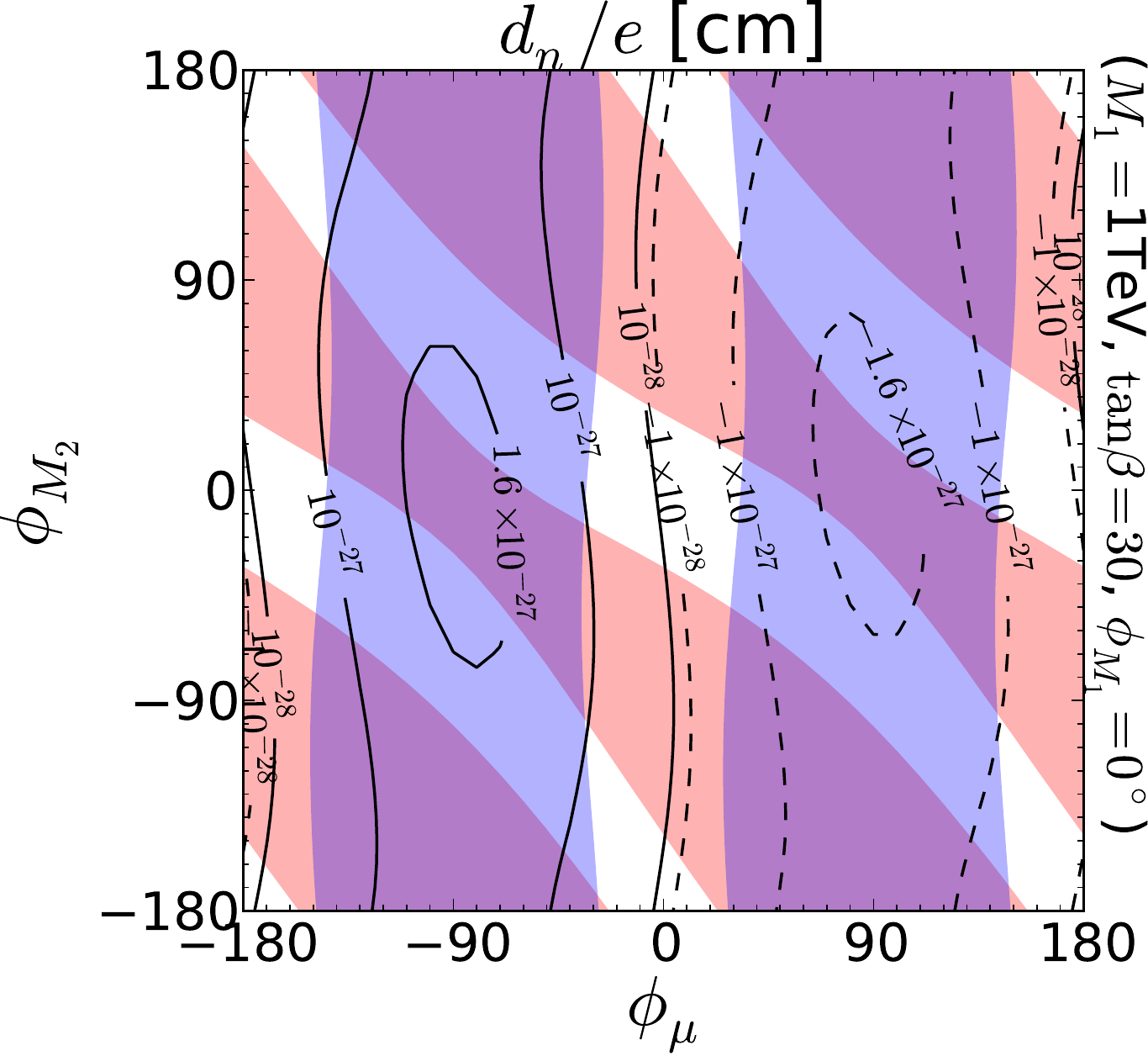}
\quad
\includegraphics[width=0.38\hsize]{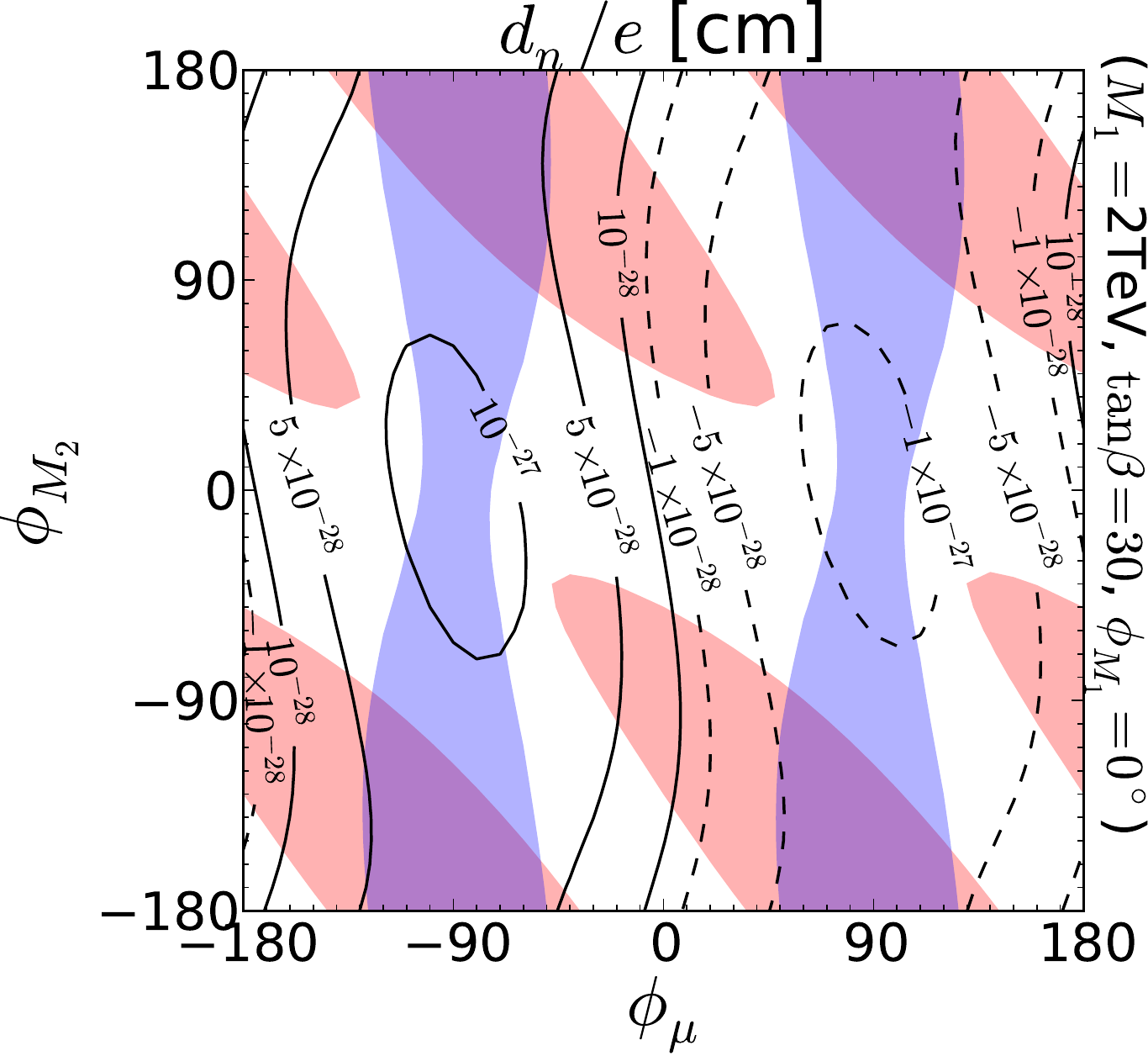}
\caption{
The EDMs for $\tan\beta=30$ and $\phi_{M_1}=0^\circ$. The left (right) panels are
for $M_1=1$~TeV ($M_1=2$~TeV).
The contours in the top, the center, and the bottom panels are those of the electron EDM,
 the mercury EDM, and the neutron EDM, respectively. 
The dashed lines show the negative values.
The red and blue shaded regions are excluded by the electron EDM and the mercury EDM, respectively.
}
\label{fig:edm_mu_vs_M2_with_fiM1-0}
\end{figure}

We discuss the $\phi_{\mu}$ and $\phi_{M_2}$ dependence of the EDMs.
The left panels in Fig.~\ref{fig:edm_mu_vs_M2_with_fiM1-0} shows the electron EDM, the mercury EDM,
and the neutron EDM with $\phi_{M_1} =0$. The shaded regions are already excluded
by the current upper bound on the EDMs. We find the combination of the electron EDM
and the mercury EDM exclude the large region of the parameter space. Both $\phi_\mu$ and
$\phi_{M_2}$ cannot be large. We also find that the electron EDM strongly
depends on $\phi_{M_2}$. 
On the other hand, $\phi_{M_2}$ dependence of the mercury EDM and the neutron EDM are milder.
We focus on $M_1$ dependence by comparing the left panels and the right panels in Fig.~\ref{fig:edm_mu_vs_M2_with_fiM1-0}
where $M_1=1$~TeV and $2$~TeV, respectively.
We find that larger $M_1$ weaken the constraint from EDM experiments
because $M_1$ is approximately the mass of the dark matter candidate here and 
heavy Higgs bosons and Higgsinos become heavier if we take larger $M_1$. 
Actually, the Higgsino mass $|\mu|$ is determined to be in $1.6$--$1.7$~TeV in order to reproduce 
the correct relic density of the dark matter, 
$\Omega_{\text{DM}}h^2=0.1198 \pm 0.0015$ for $M_1=1$~TeV.
For larger $M_1$ such as $M_1=2$~TeV, the Higgsino mass becomes larger as $|\mu|\sim 2.3$~TeV.
In Tab.~\ref{tab:spectrum}, we show the mass spectrum of SUSY particles and the extra Higgs bosons 
in our benchmark points for the cases of $M_1=1$~TeV and $M_1=2$~TeV.
\begin{table}
\caption{The mass spectrum of the SUSY particles and extra Higgs bosons in our scenario.\label{tab:spectrum}}	
\begin{tabular}{c|c|c|c|c|c|c}
Cases&$m_{\tilde{\chi}^0_1}$&$m_{H^{\pm}}\simeq m_{H,A}$&
	$m_{\tilde{\chi}^0_{2,3}}\simeq m_{\tilde{\chi}^{\pm}_1}$&$m_{\tilde{\chi}^0_4}\simeq m_{\tilde{\chi}^{\pm}_2}\simeq m_{\tilde{g}}$&$m_{\tilde{t}_{1,2}}\simeq m_{\tilde{b}_1}$&Other sfermions\\ \hline
$M_1=1$~TeV&$\sim 1$~TeV&$\sim 2$~TeV&$1.6$ -- $1.7$~TeV&$\sim 10$~TeV&$\sim 7$~TeV&$\sim 100$~TeV\\ \hline
$M_1=2$~TeV&$\sim 2$~TeV&$\sim 4$~TeV&$\sim 2.3$~TeV&$\sim 10$~TeV&$\sim 7$~TeV&$\sim 100$~TeV\\ \hline
\end{tabular}
\end{table}

\begin{figure}[tb]
\includegraphics[width=0.38\hsize]{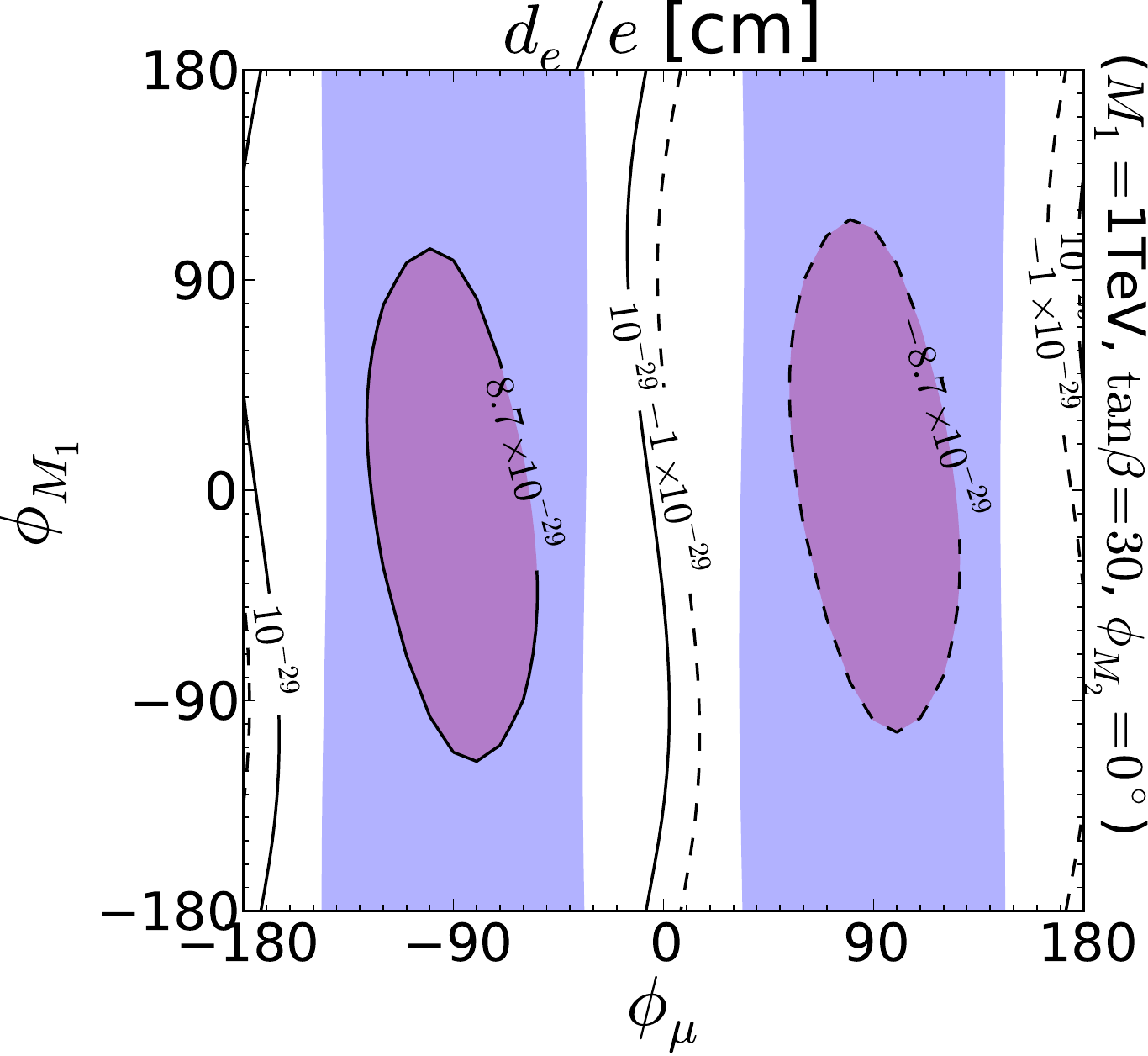}
\qquad
\includegraphics[width=0.38\hsize]{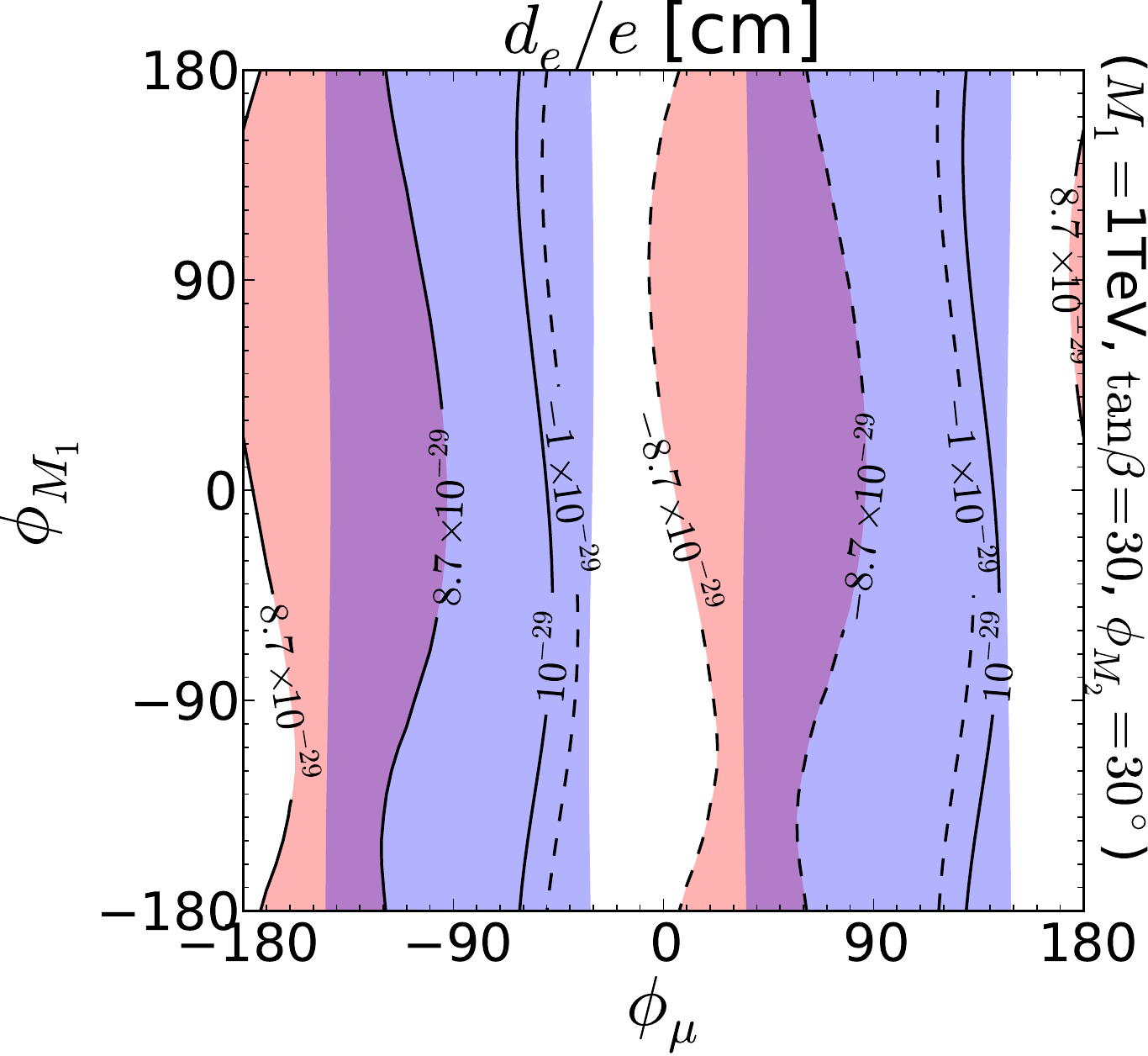}
\\
\includegraphics[width=0.38\hsize]{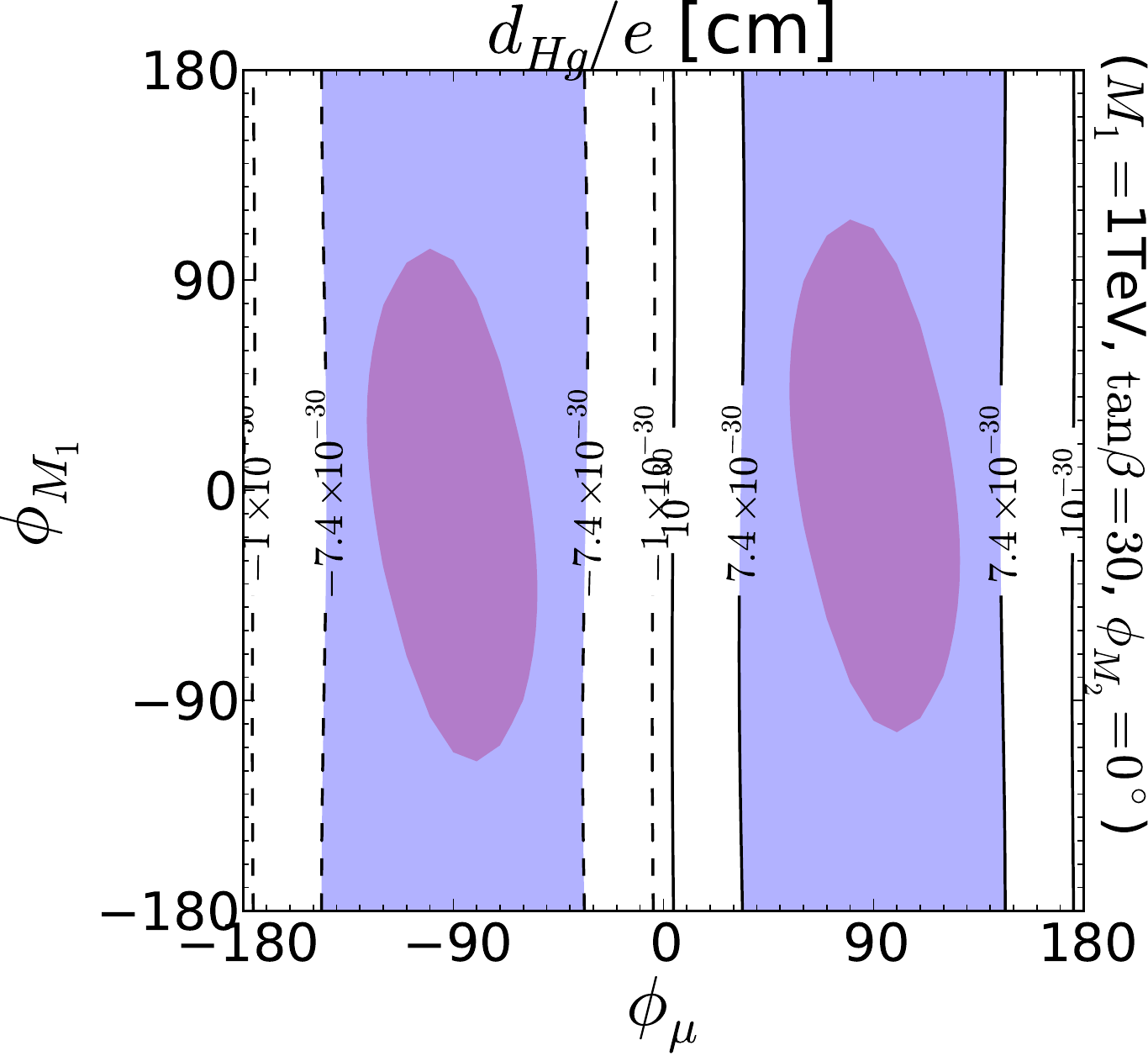}
\qquad
\includegraphics[width=0.38\hsize]{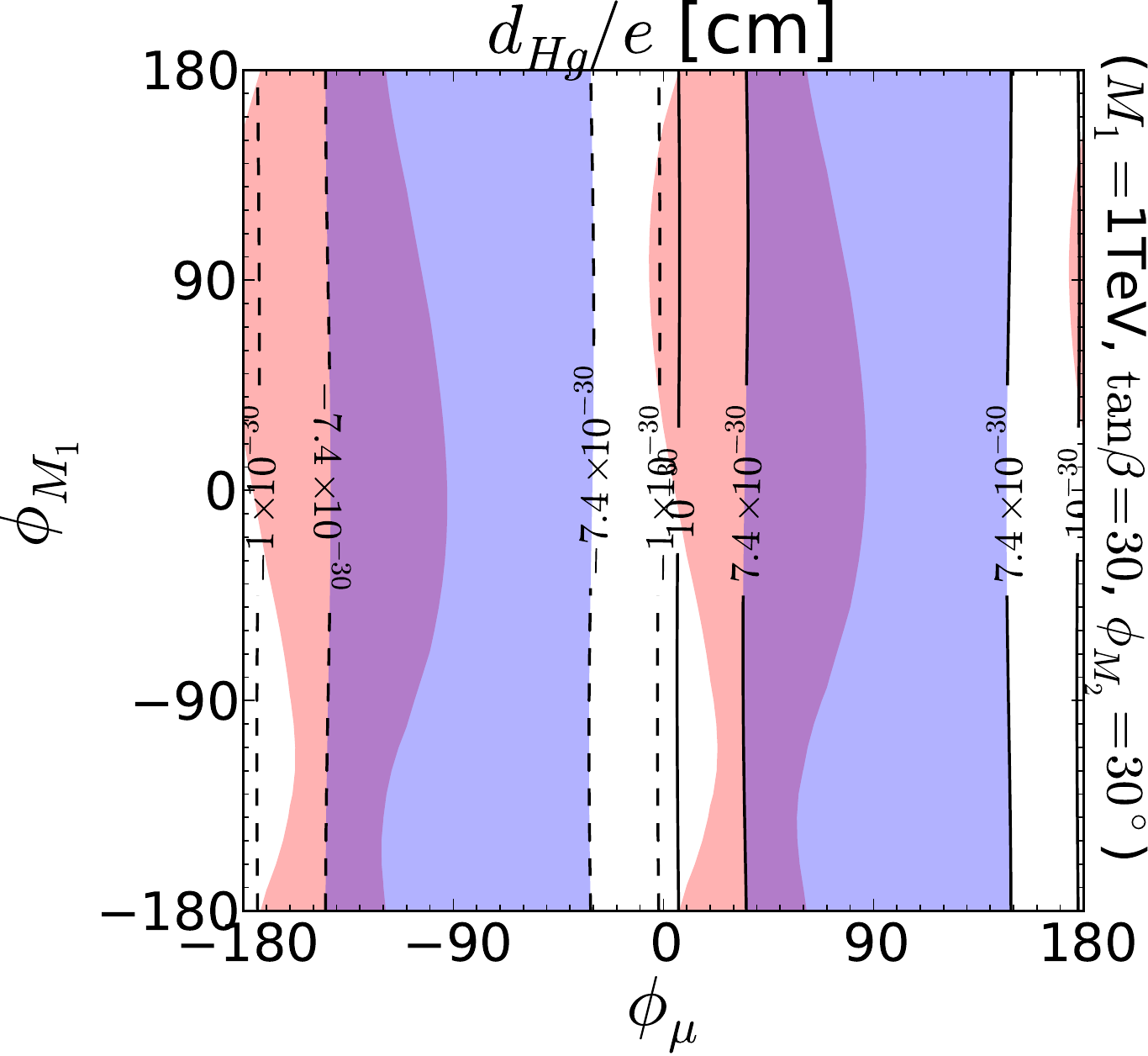}
\\
\includegraphics[width=0.38\hsize]{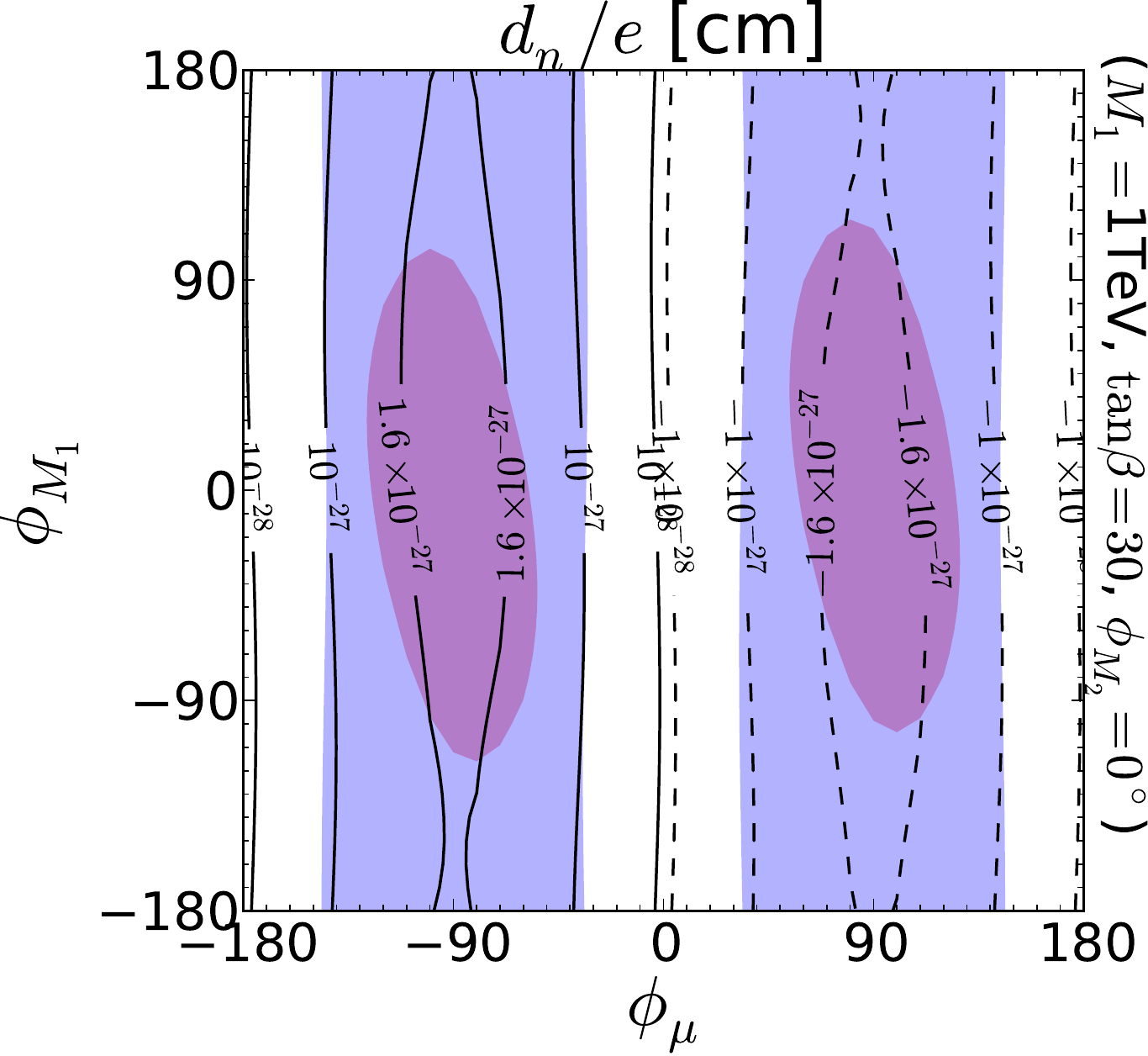}
\qquad
\includegraphics[width=0.38\hsize]{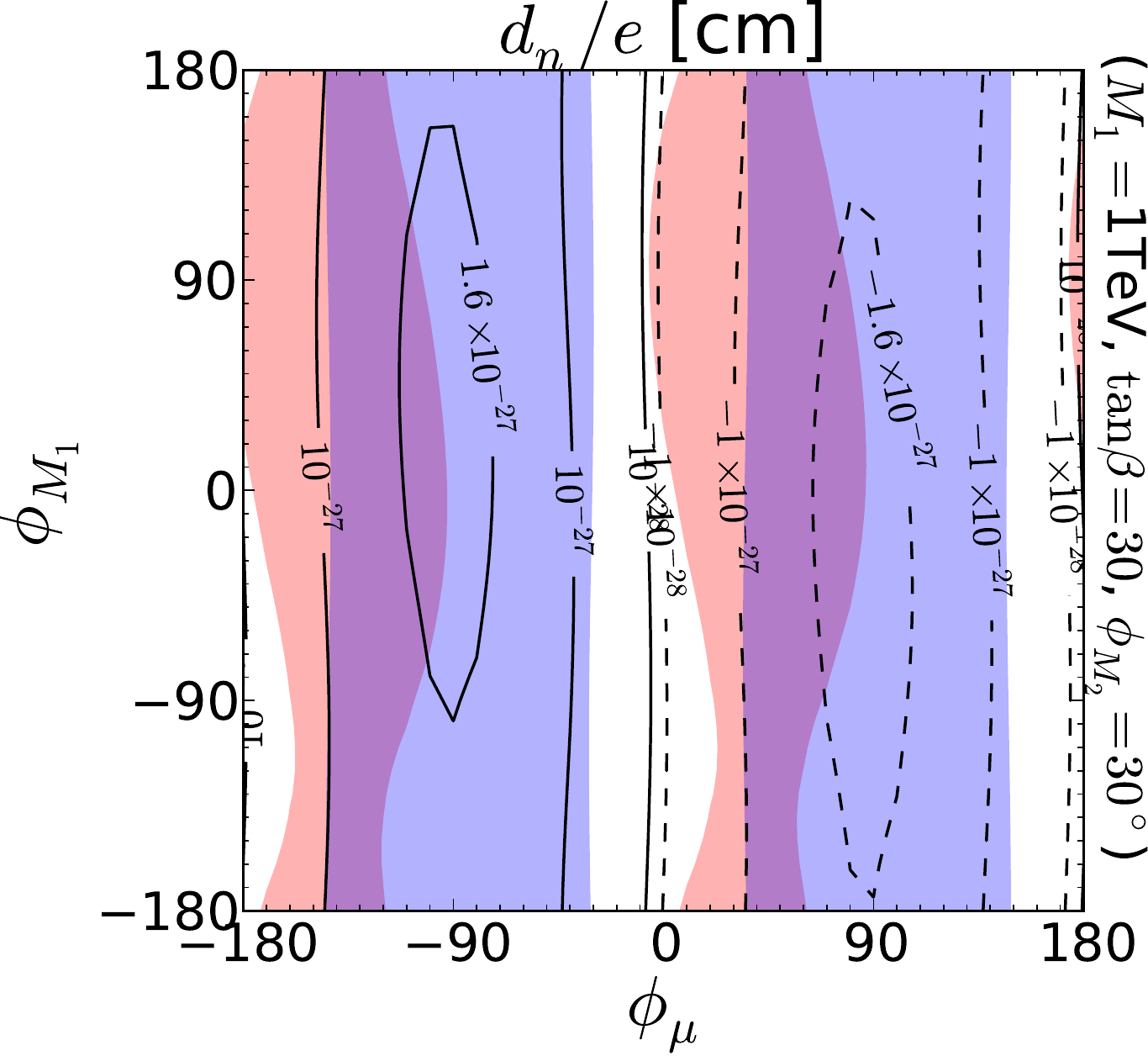}
\caption{
The EDMs for $M_1=1$~TeV and $\tan\beta=30$.
In the left (right) panels, $\phi_{M_2}=0^\circ$ ($30^\circ$). 
The shadings and contours are the same as in Fig.~\ref{fig:edm_mu_vs_M2_with_fiM1-0}.
}
\label{fig:M1-dependence}
\end{figure}

The $\phi_{M_1}$ dependence can be seen in Fig.~\ref{fig:M1-dependence}.
After taking into account the constraint from the mercury EDM, 
we find that the mercury EDM and the neutron EDM are almost independent of $\phi_{M_1}$.
On the other hand, the electron EDM has mild but visible dependence on $\phi_{M_1}$.
Thus the electron EDM is important for determining $\phi_{M_1}$.

Most of the parameter space in Figs.~\ref{fig:edm_mu_vs_M2_with_fiM1-0} and \ref{fig:M1-dependence}
are within the future prospects of the electron EDM and the neutron EDM shown in Eqs.~\eqref{eq:prospect_de} and \eqref{eq:prospect_dn}. 
The neutron and the mercury EDMs are sensitive to $\phi_{\mu}$, and also weakly depend on $\phi_{M_2}$. 
On the other hand, the electron EDM is sensitive to $\phi_{M_2} + \phi_{\mu}$, and weakly depend on $\phi_{M_1}$. 
The correlation among the EDMs in future experiments provide a strong clue to 
explore the CP violation in the SUSY breaking sector.
In our scenario,  we can determine the imaginary part of SUSY breaking parameters such as 
$\mathrm{Im}(M_2)$. If the absolute values of those parameters
are determined by the direct 
measurement of SUSY particle masses, we can determine the CP phases.

\begin{figure}[tb]
\includegraphics[width=0.38\hsize]{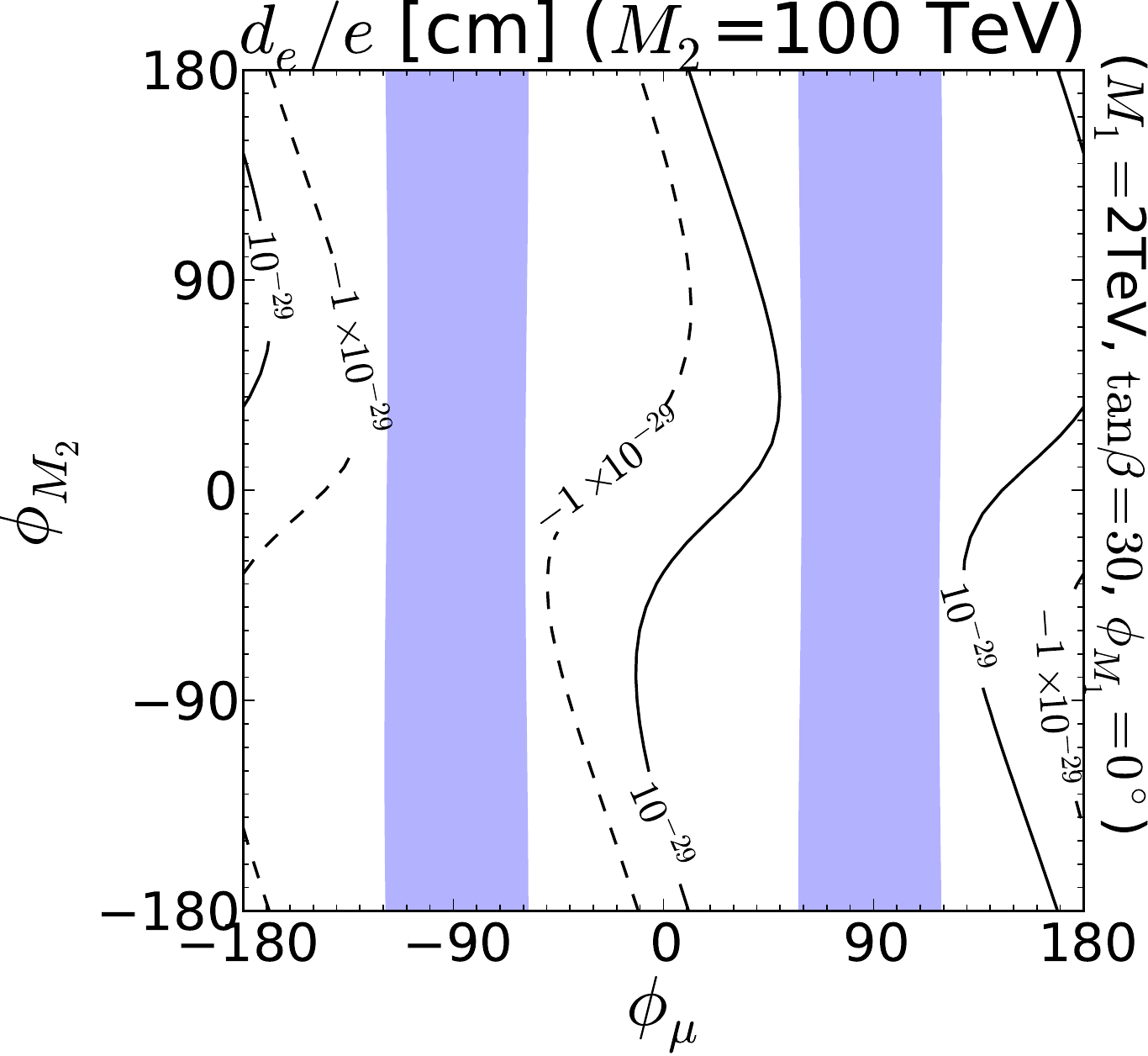}
\quad
\includegraphics[width=0.38\hsize]{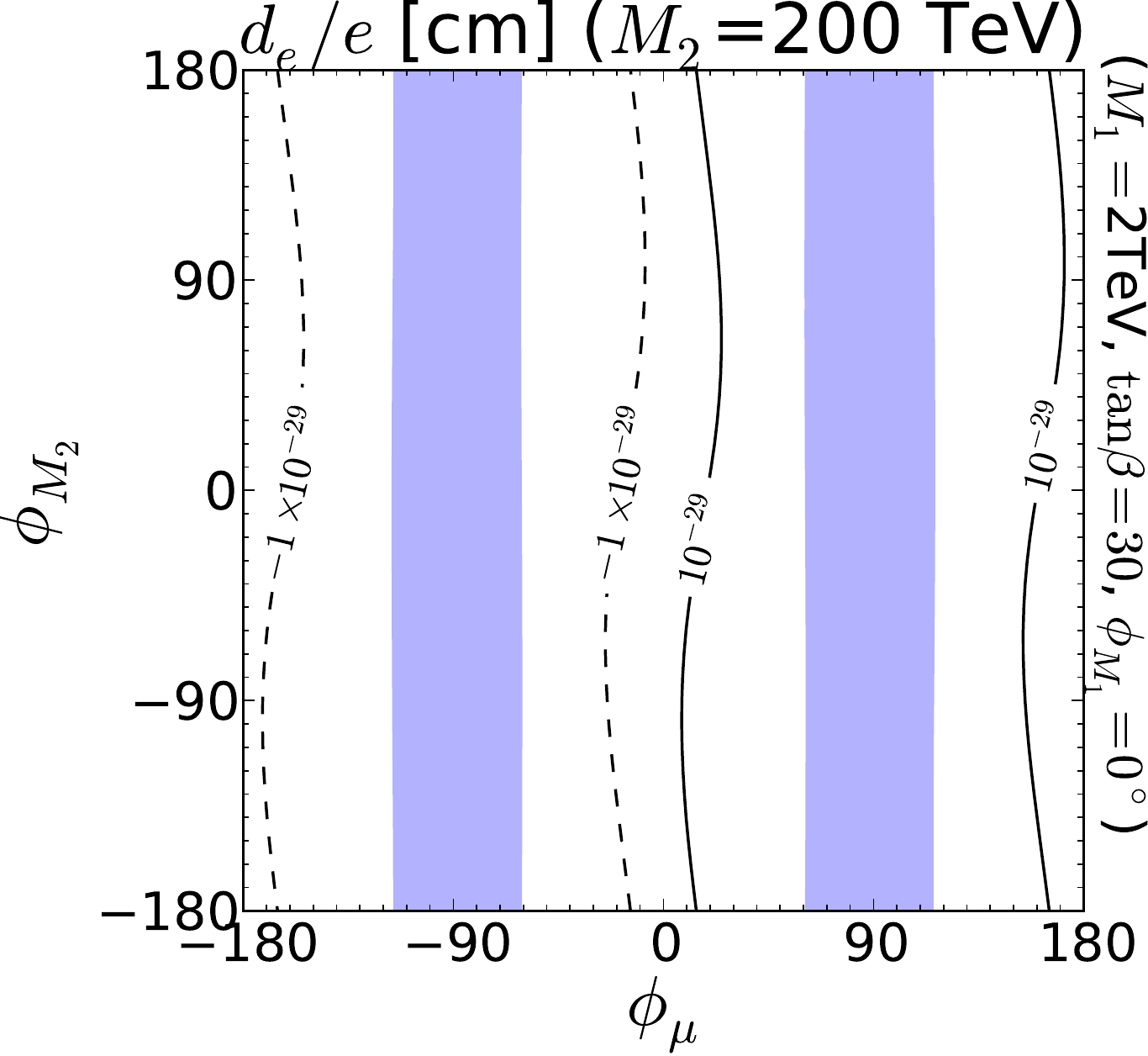}
\\
\includegraphics[width=0.38\hsize]{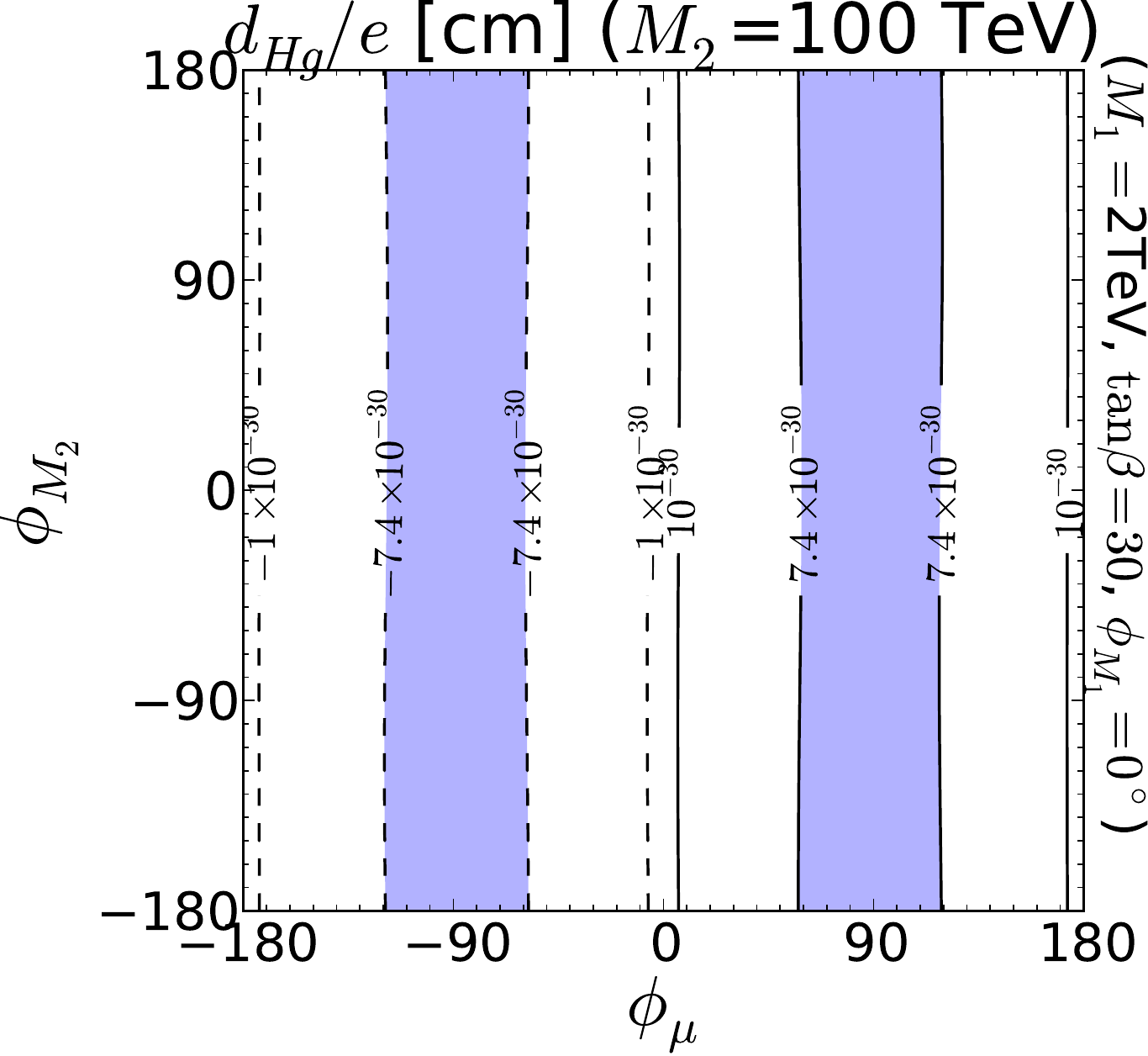}
\quad
\includegraphics[width=0.38\hsize]{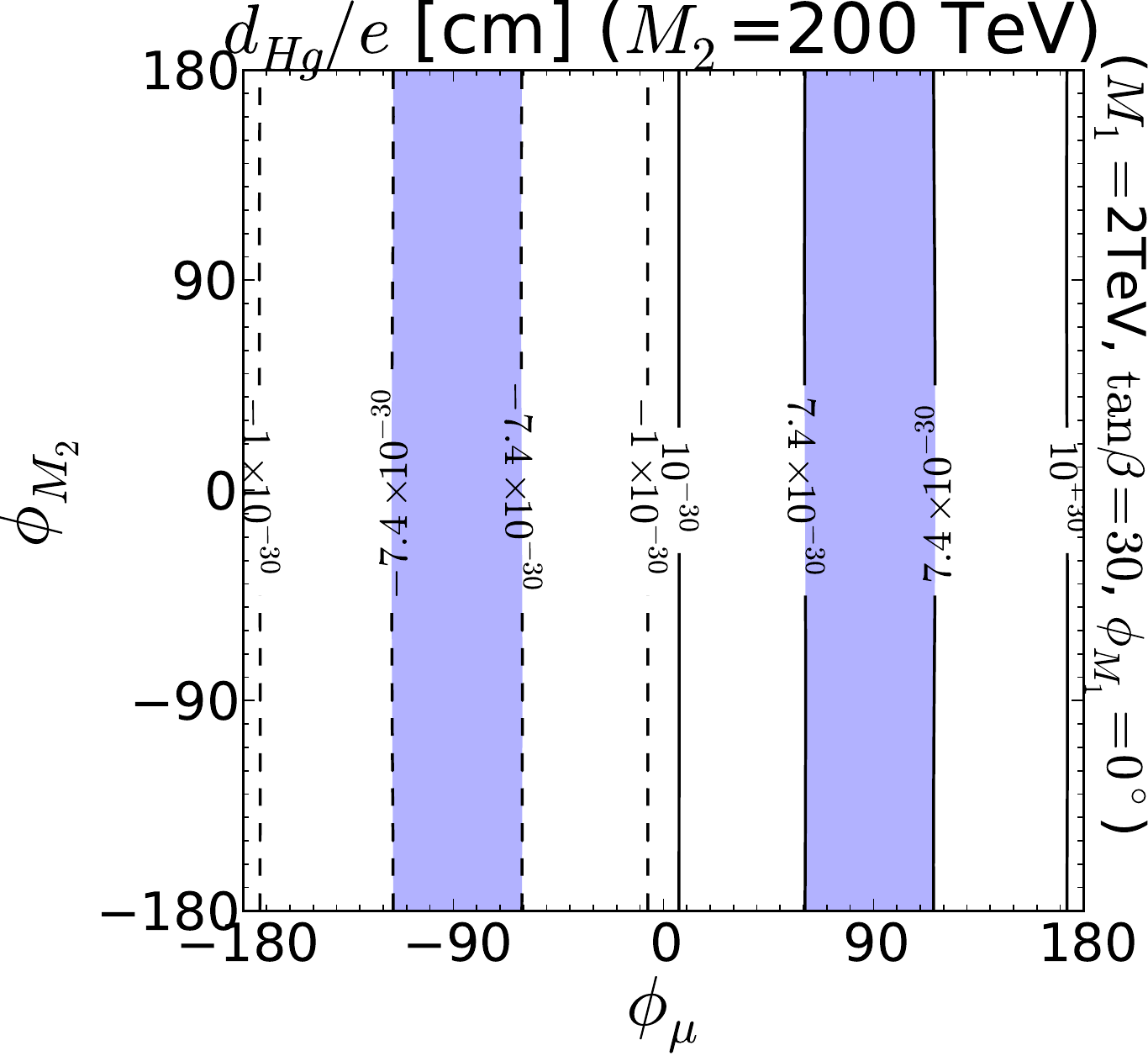}
\\
\includegraphics[width=0.38\hsize]{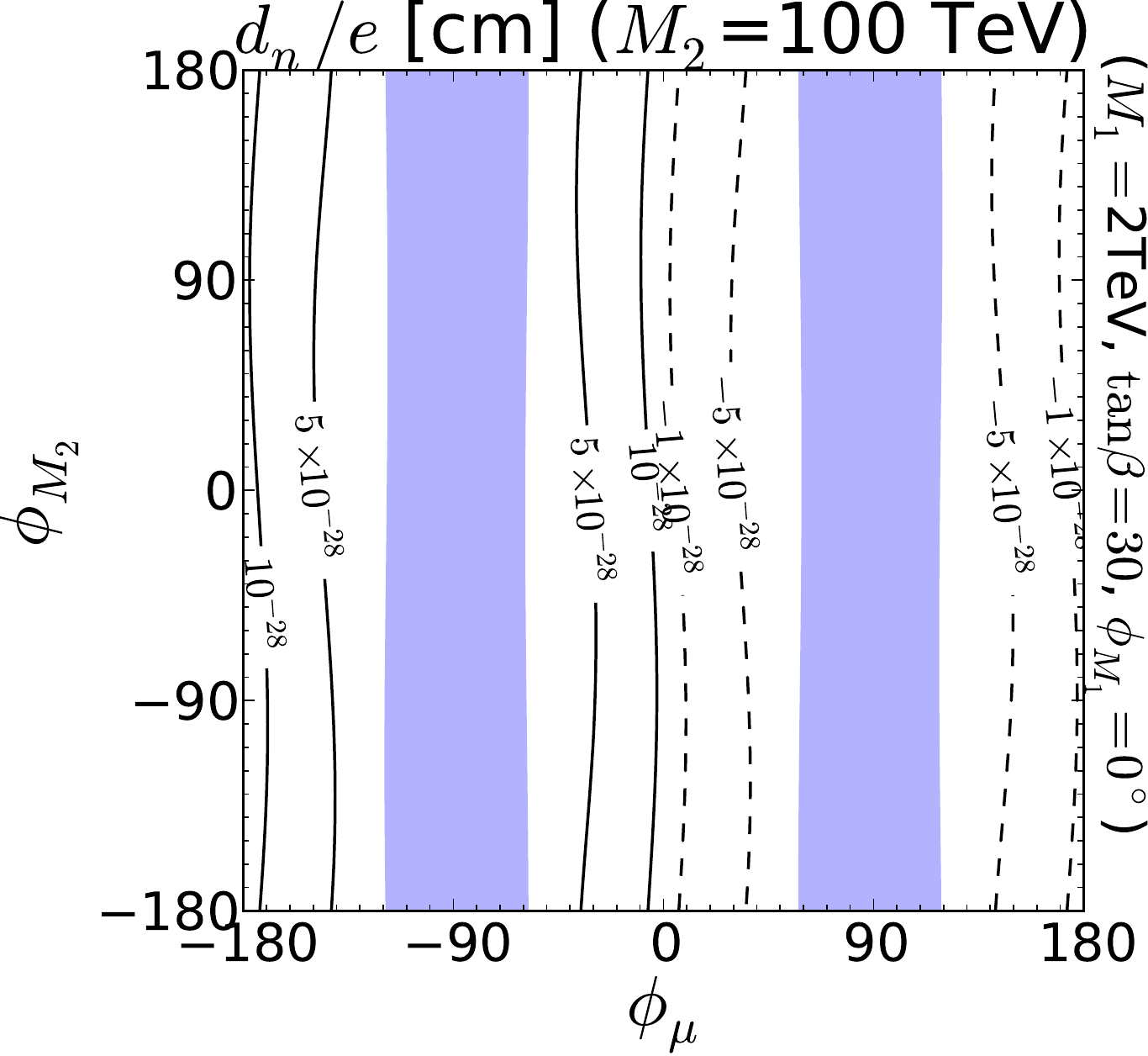}
\quad
\includegraphics[width=0.38\hsize]{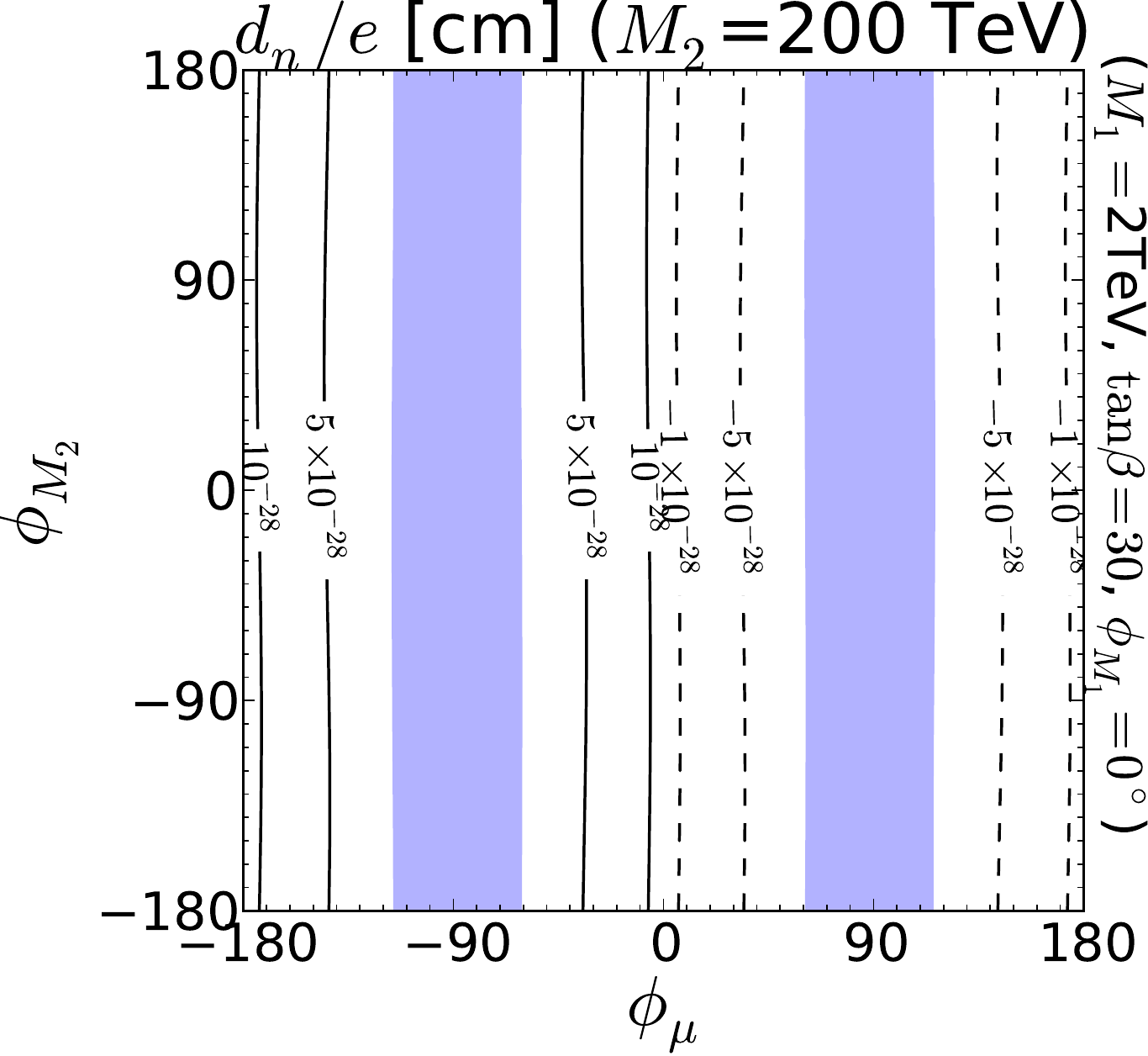}
\caption{
The EDMs for $M_1=2$~TeV, $\tan\beta=30$, and $\phi_{M_2}=0^\circ$
In the left (right) panels, $M_2=100$~TeV ($200$~TeV). 
The shadings and contours are the same as in Fig.~\ref{fig:edm_mu_vs_M2_with_fiM1-0}.
Notice that no red colored region in this parameter space.
}
\label{fig:M2-100}
\end{figure}

\begin{figure}[tb]
\includegraphics[width=0.49\hsize]{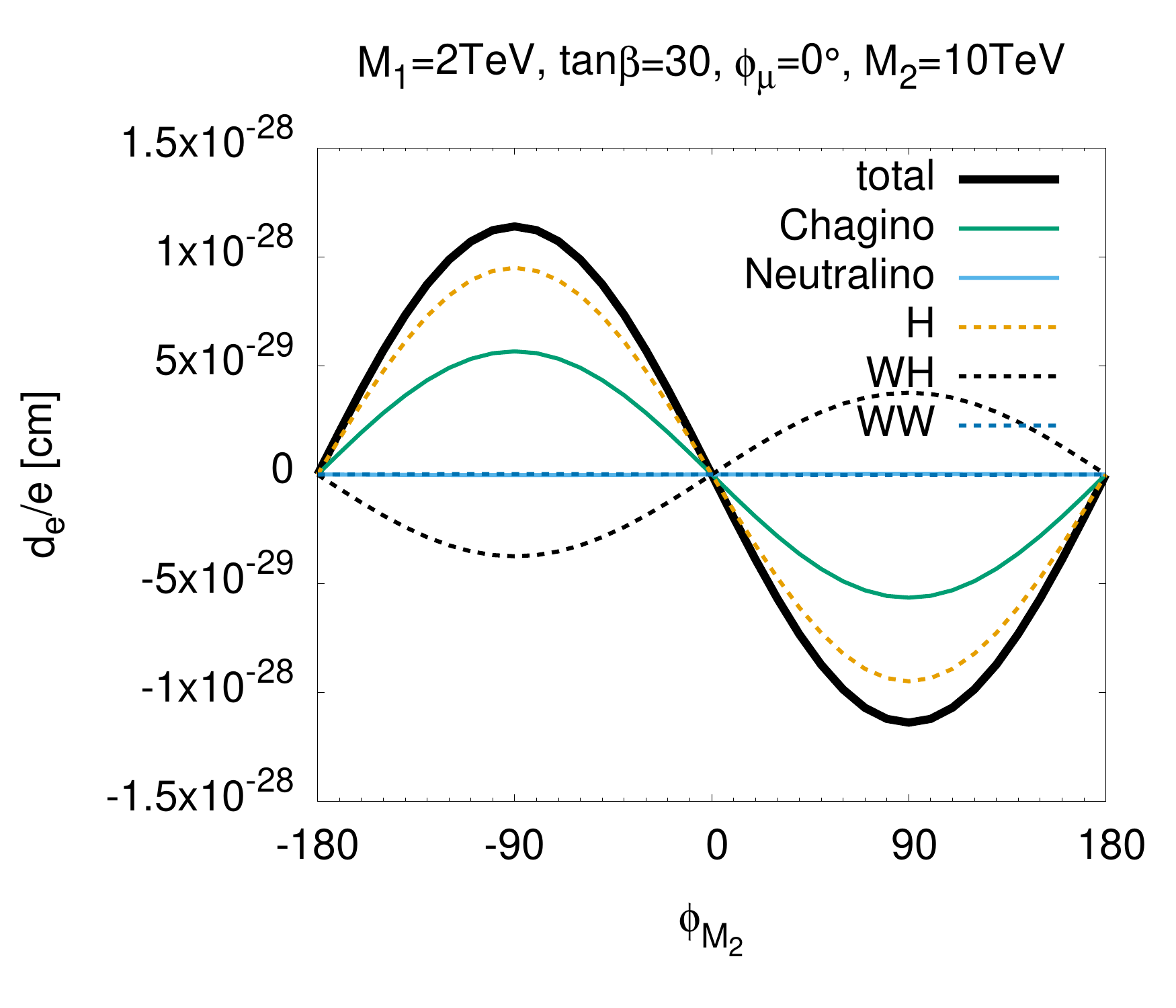}
\includegraphics[width=0.49\hsize]{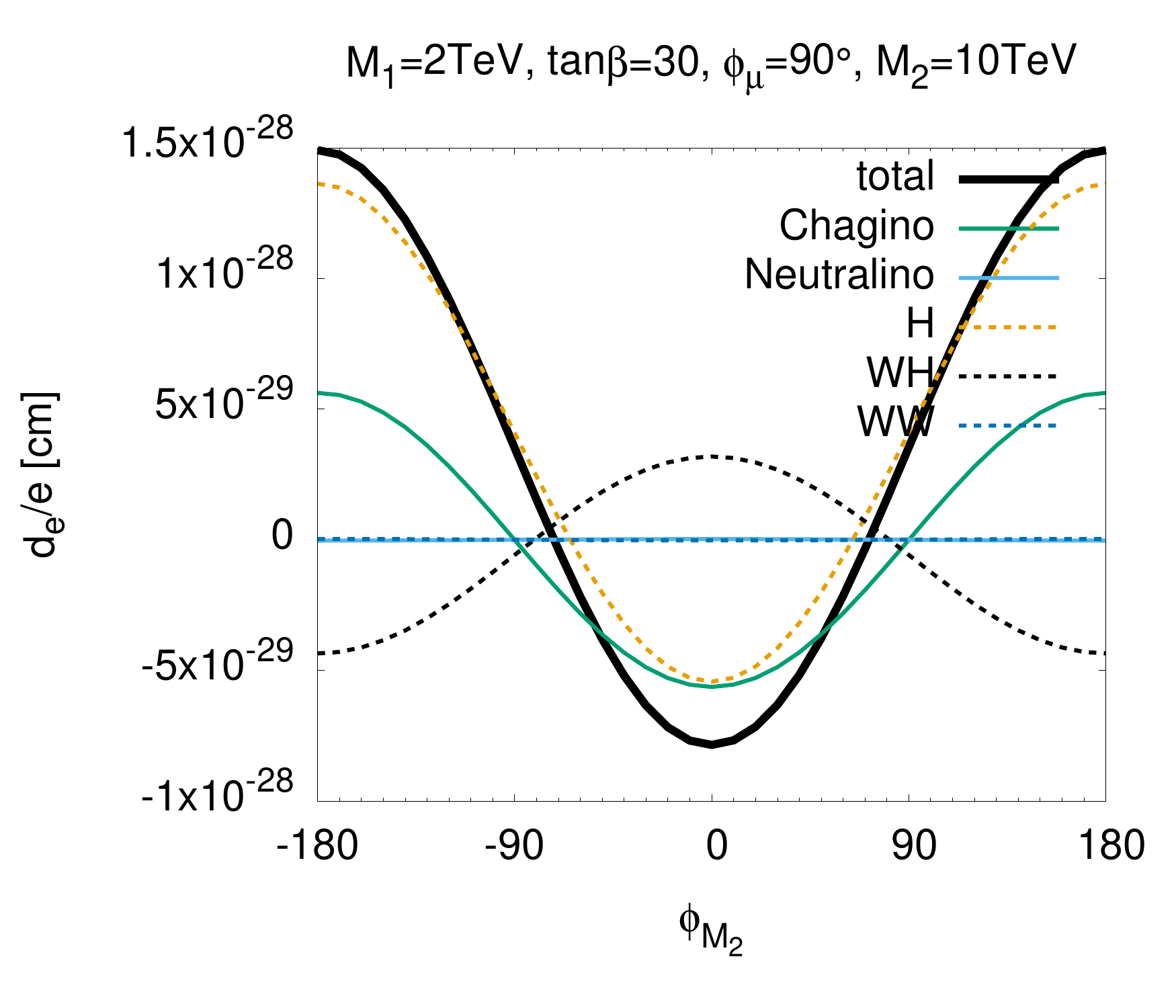}
\includegraphics[width=0.49\hsize]{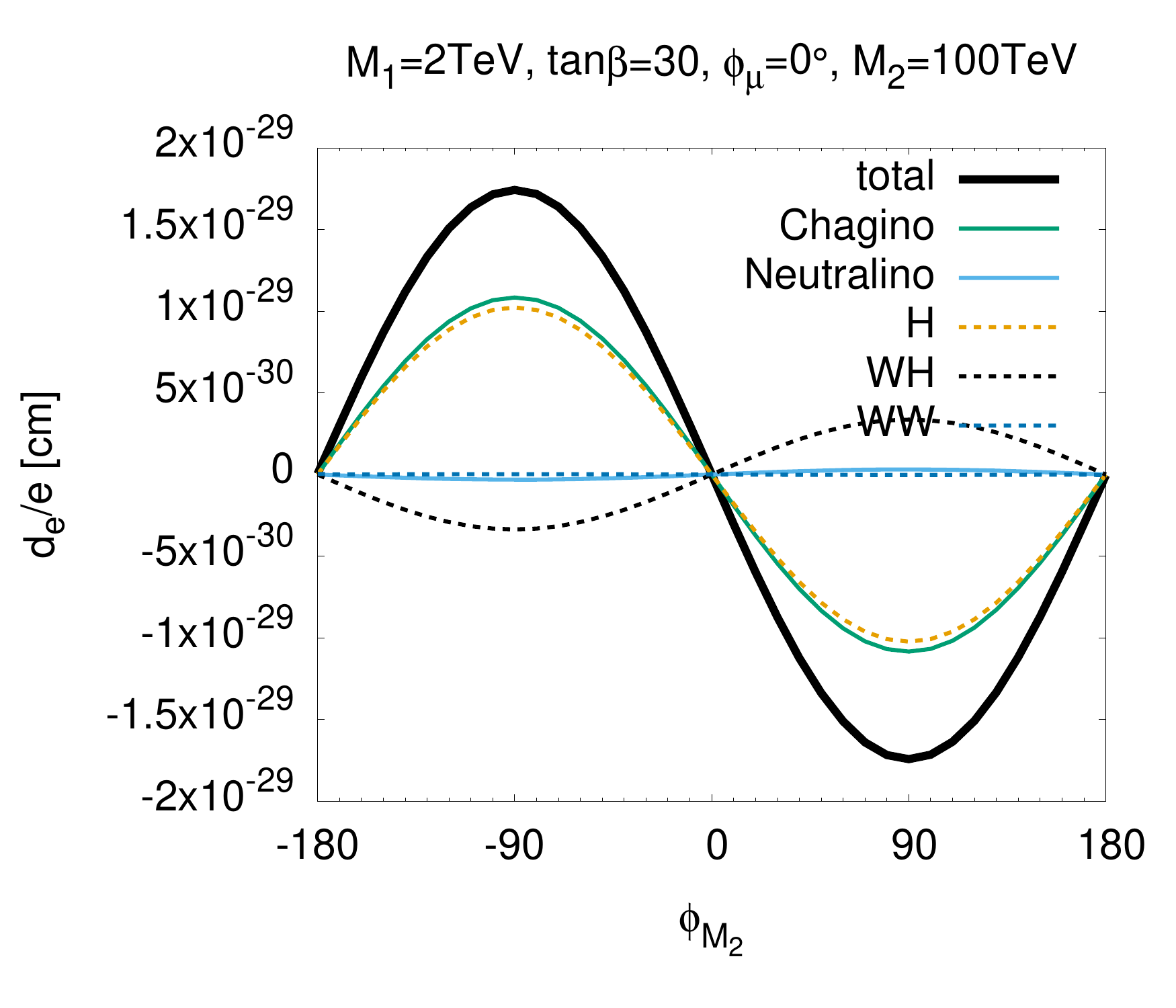}
\includegraphics[width=0.49\hsize]{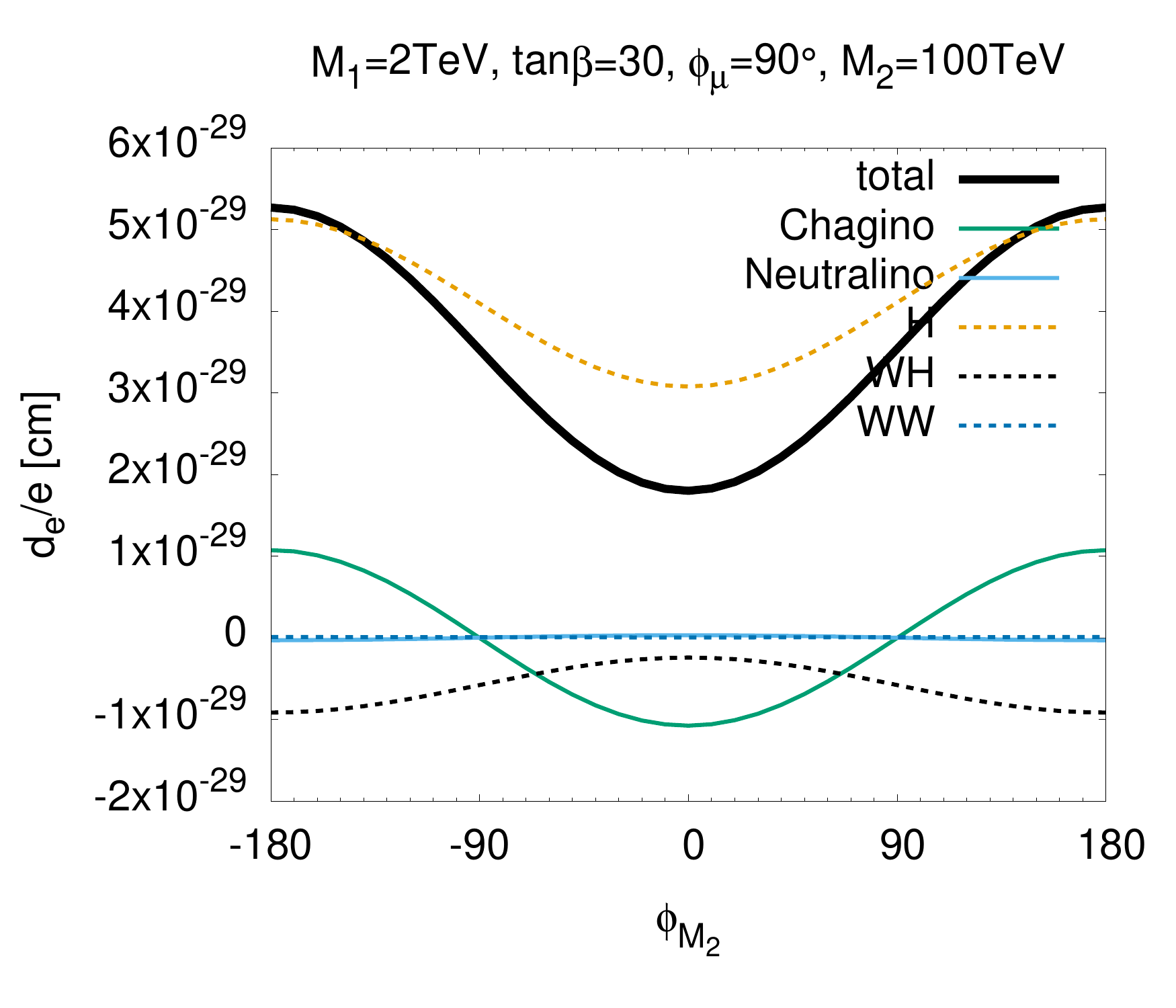}
\includegraphics[width=0.49\hsize]{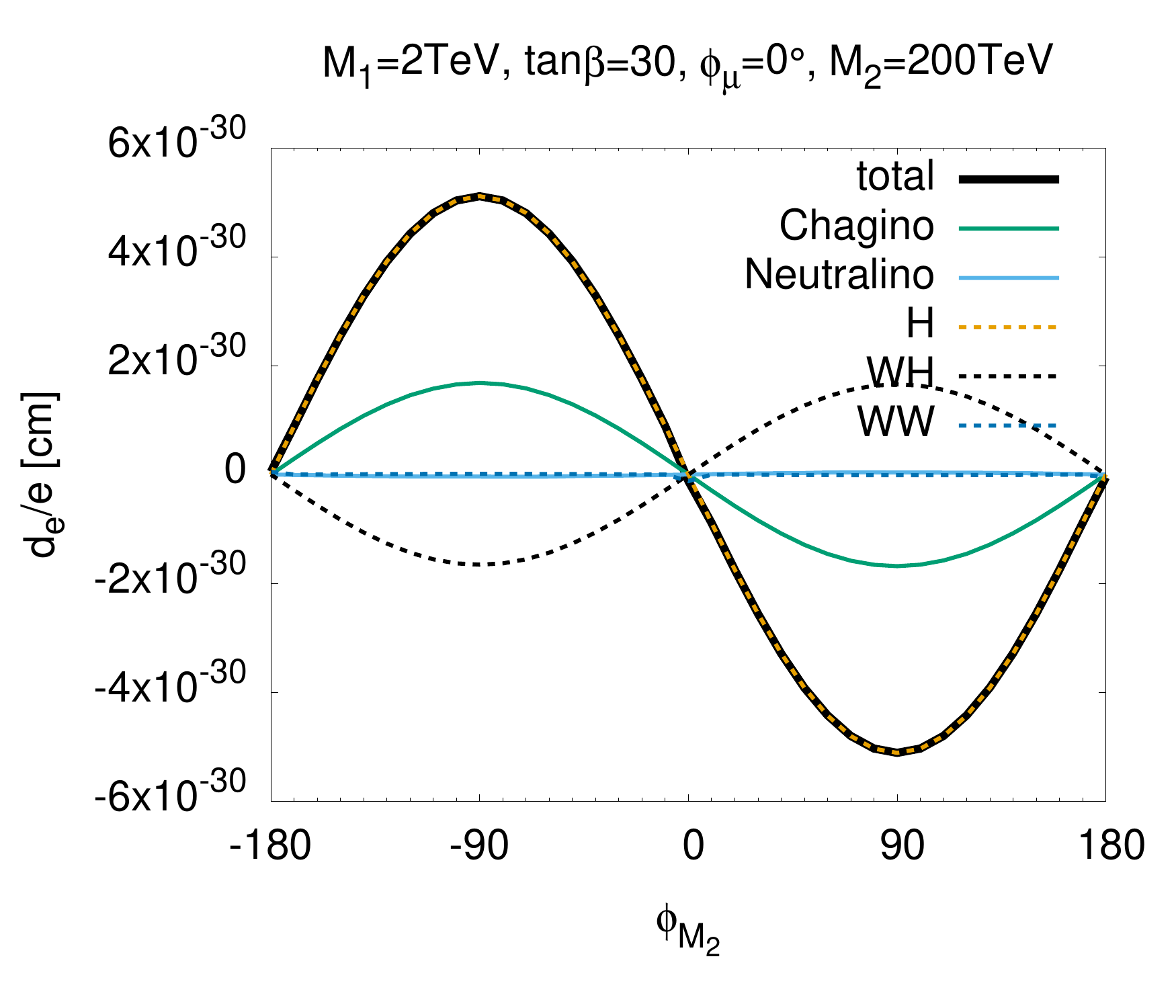}
\includegraphics[width=0.49\hsize]{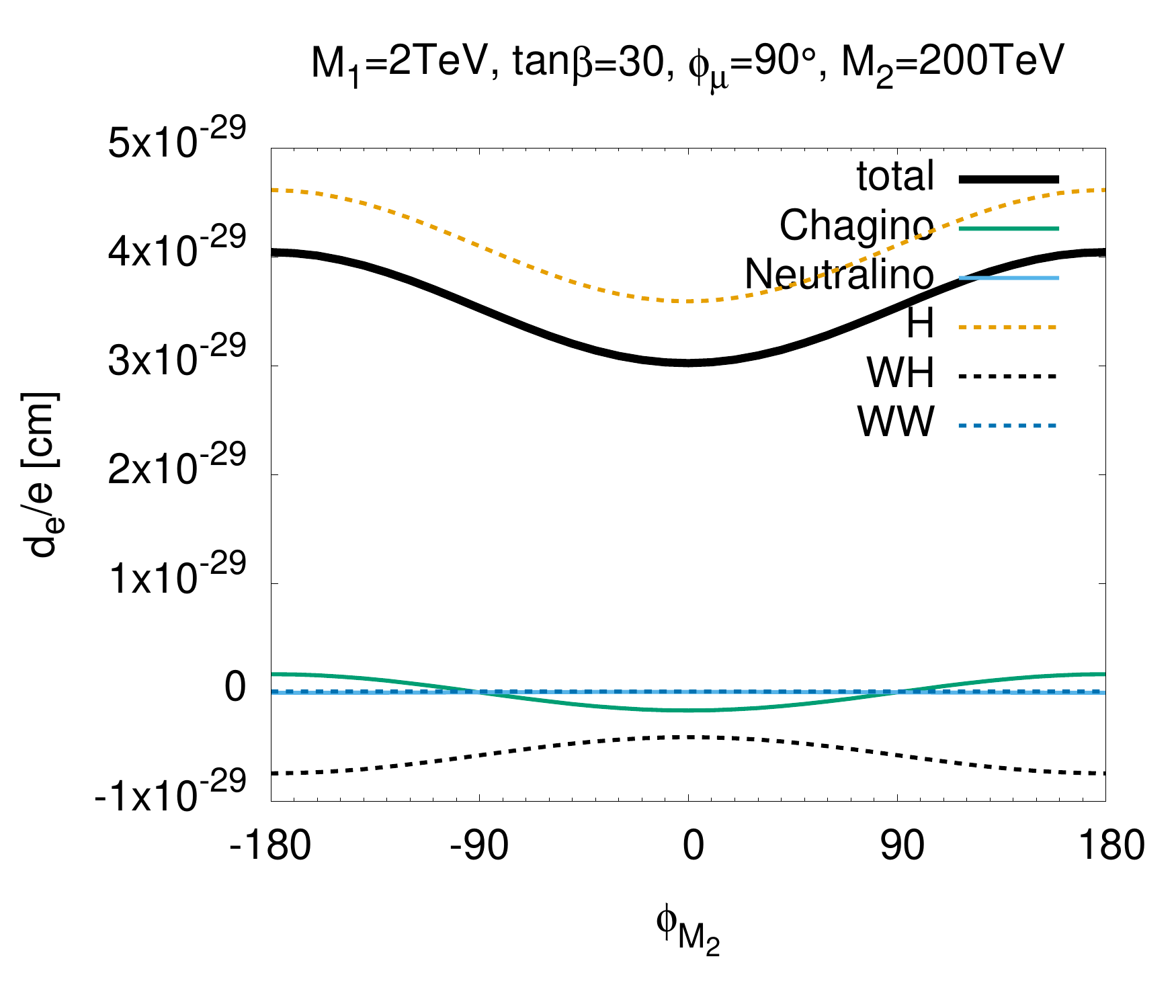}
\caption{
Contributions to the electron EDM from each diagram.
The solid green and blue curved lines are 
the contributions from the one-loop diagrams with chargino and neutralino, respectively.
The dashed orange, dashed black, and dashed blue curved lines are the contributions
from the Barr-Zee diagrams with $H$-$\gamma$-$\gamma$, $W^{\pm}$-$H^{\mp}$-$\gamma$, 
and $W^{+}$-$W^{-}$-$\gamma$ effective vertices, respectively, where $H$ denotes three neutral Higgs bosons.
}
\label{fig:anatomy-de}
\end{figure}

\begin{figure}[tb]
\includegraphics[width=0.49\hsize]{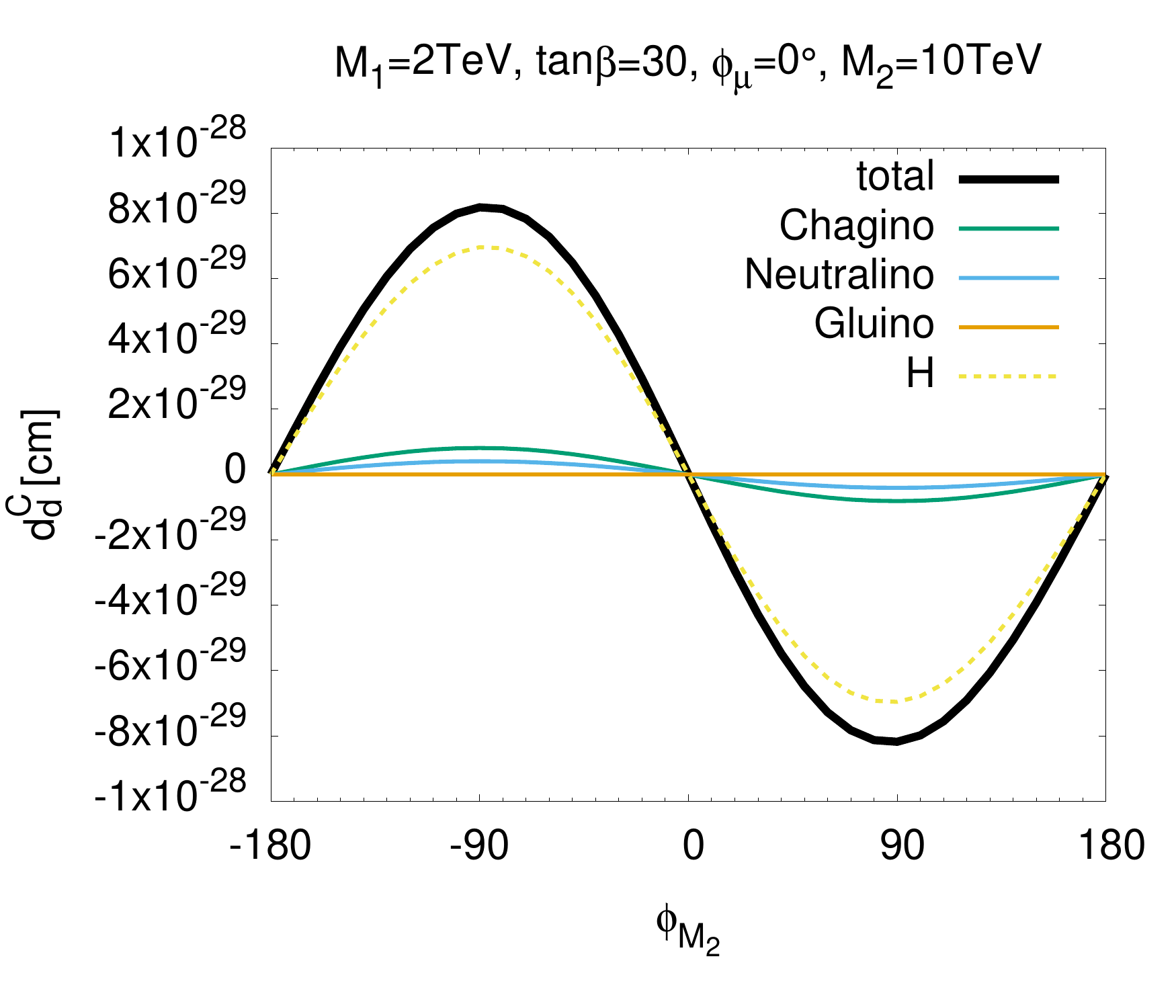}
\includegraphics[width=0.49\hsize]{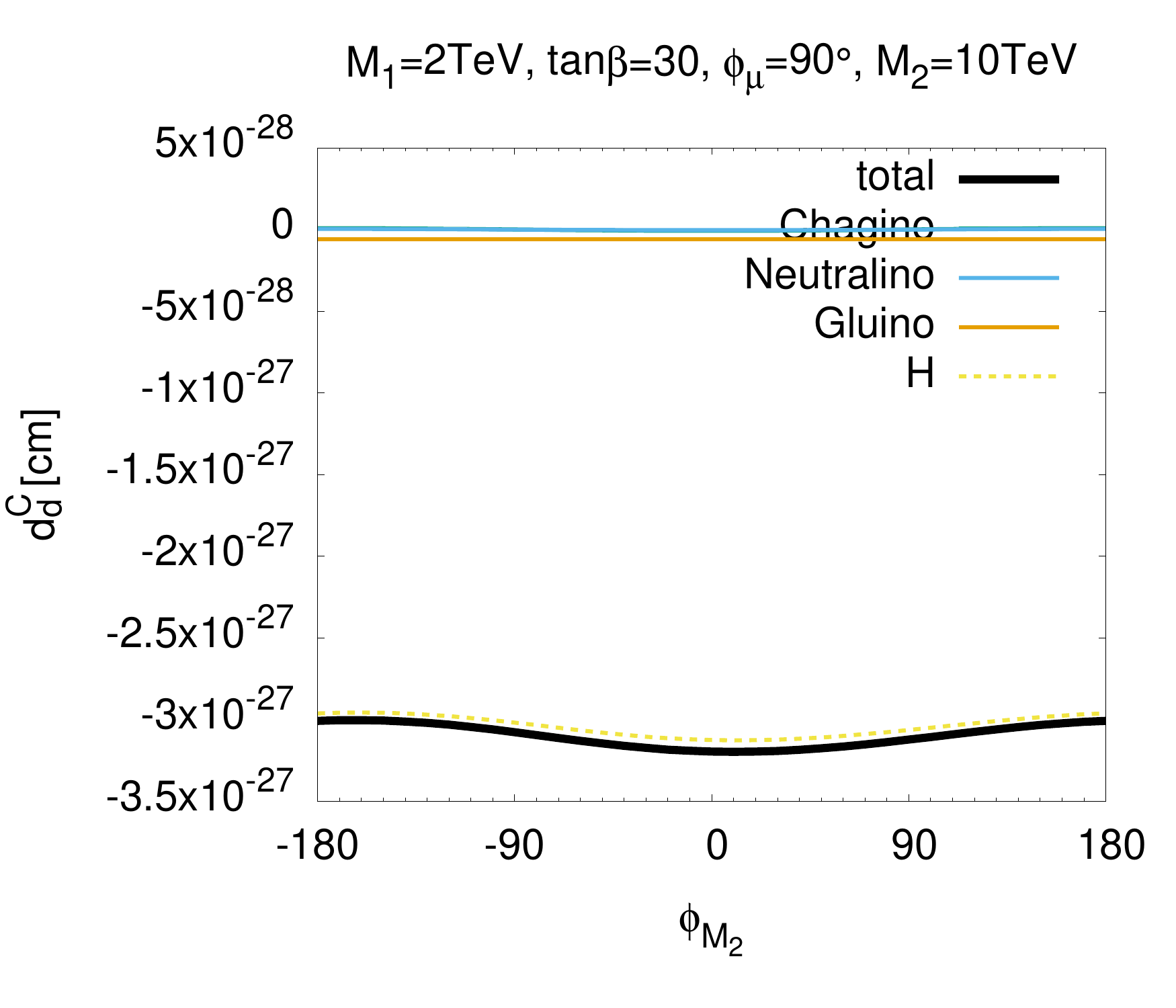}
\includegraphics[width=0.49\hsize]{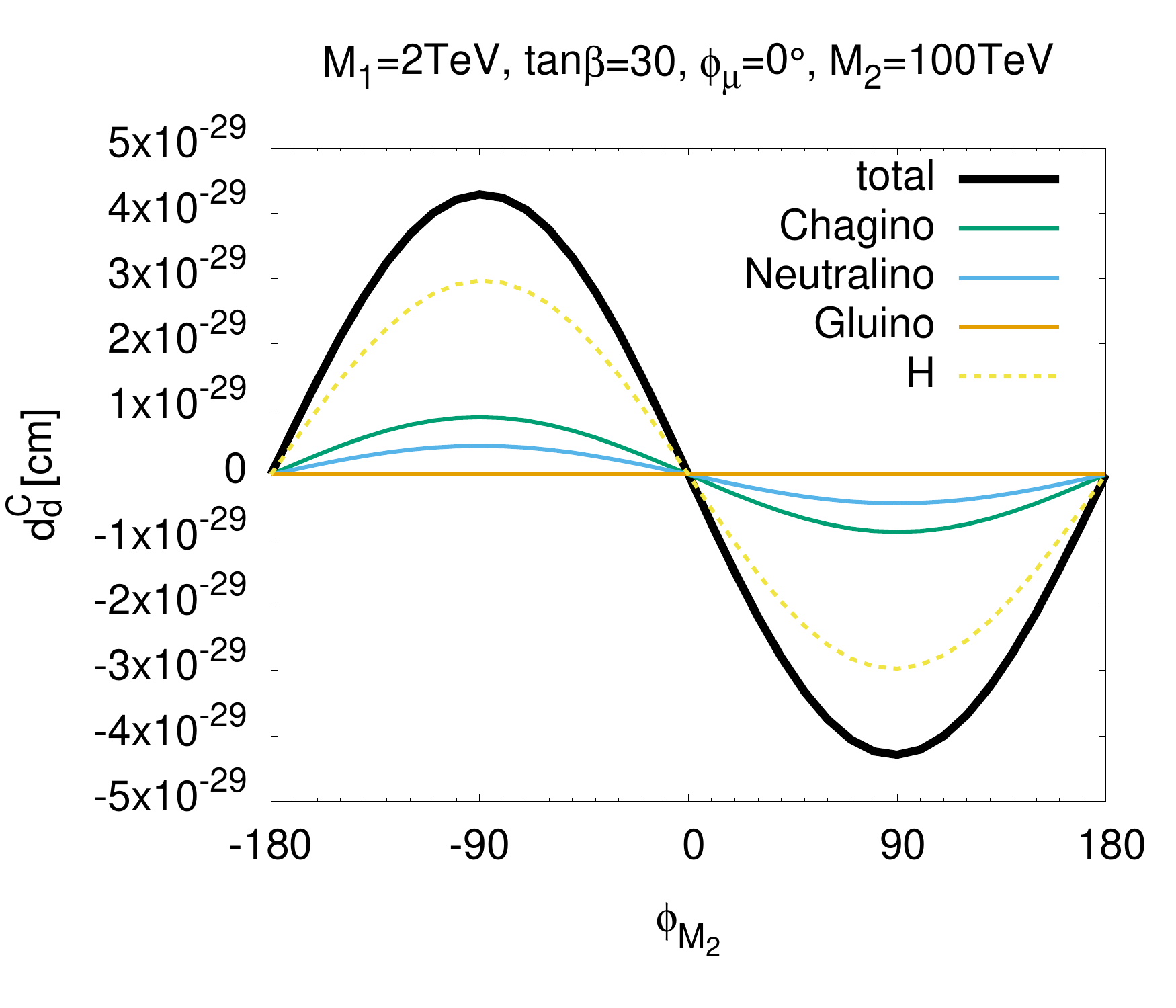}
\includegraphics[width=0.49\hsize]{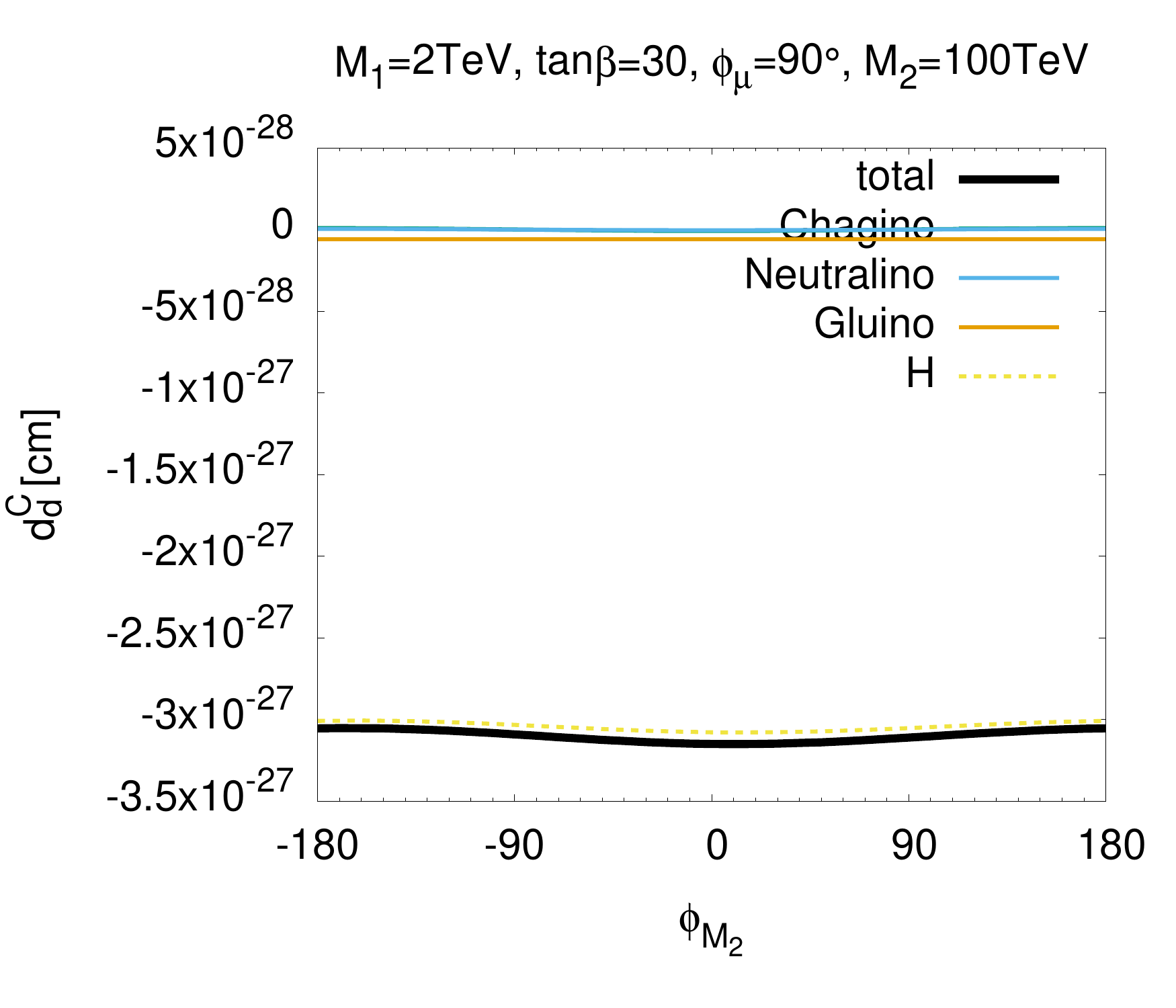}
\includegraphics[width=0.49\hsize]{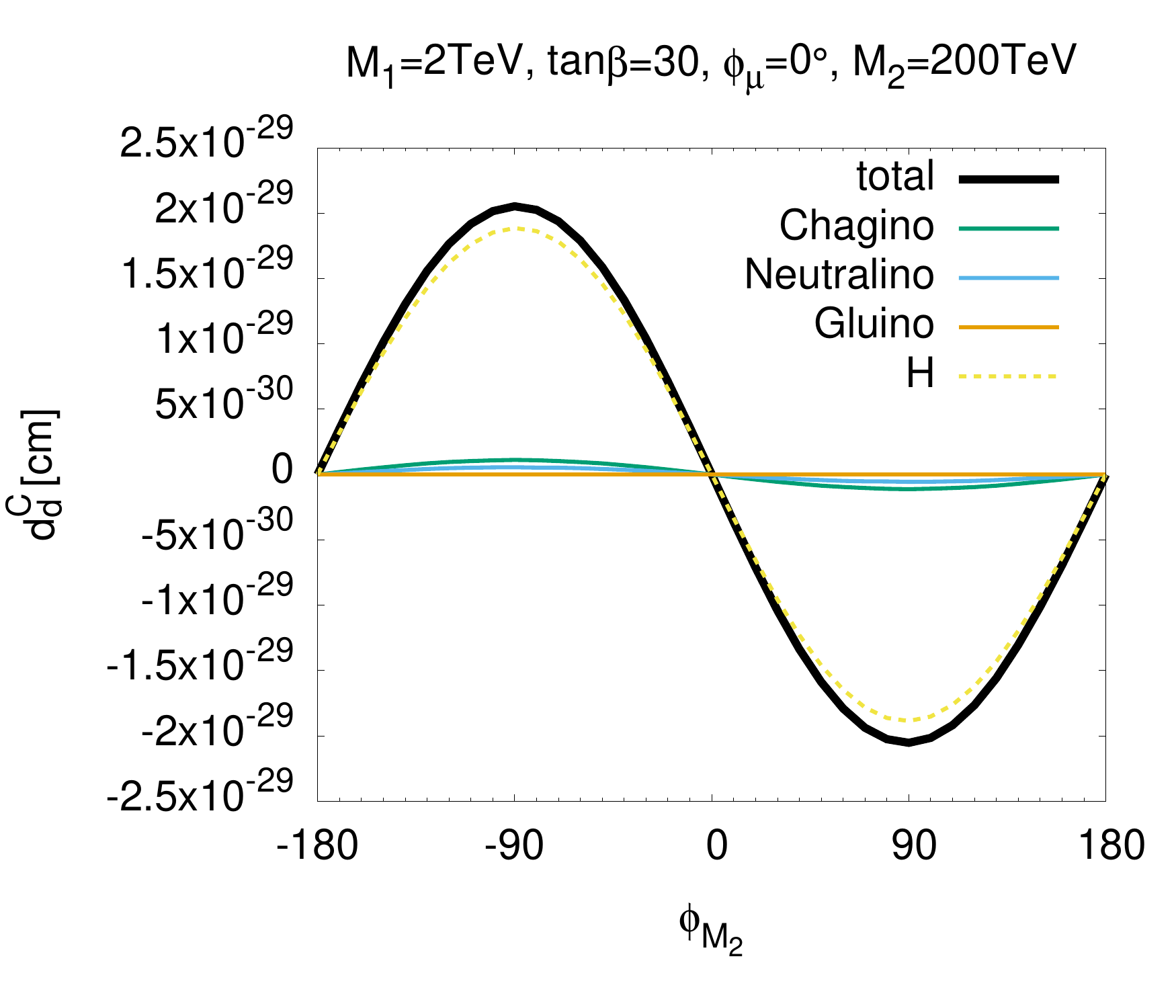}
\includegraphics[width=0.49\hsize]{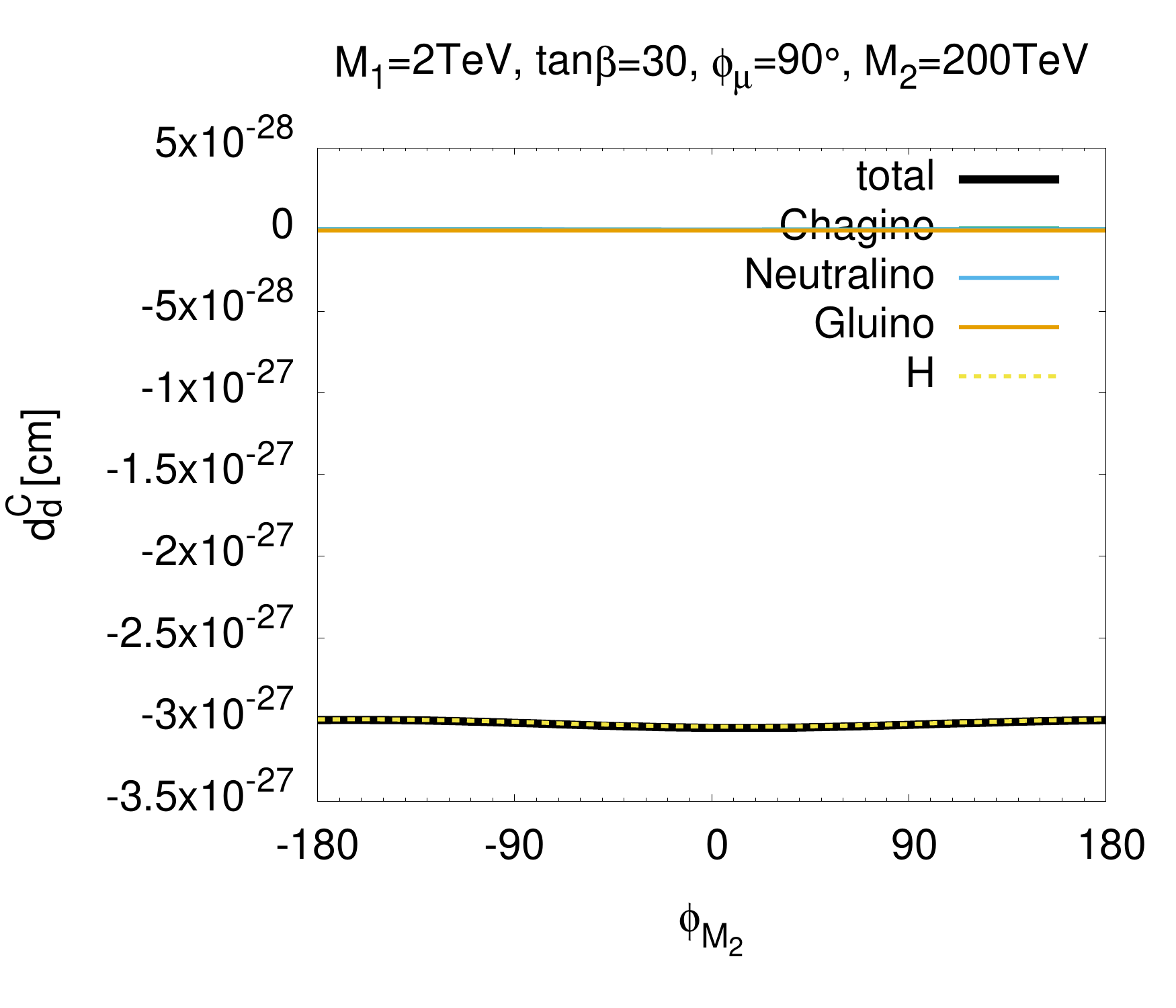}
\caption{
Contributions to the down quark chromo-EDM from each diagram.
The solid green, blue, and orange curved lines are 
the contributions from the one-loop diagrams with chargino, neutralino, and gluino, respectively.
The dashed yellow curved lines are the contributions
from the Barr-Zee diagrams with $H$-$g$-$g$.
}
\label{fig:anatomy-dcd}
\end{figure}

We next discuss the decoupling property for larger $M_2$.
Figure~\ref{fig:M2-100} shows the EDMs for $M_2=100$~TeV and $200$~TeV.
Comparing with the right panels in Fig.~\ref{fig:edm_mu_vs_M2_with_fiM1-0}, 
we find that larger $M_2$ drastically reduce the values of the EDMs.
In particular, the electron EDM is sensitive to the $M_2$ choice
but the mercury EDM is not. 
This property can be understood as follows. 
The dominant contribution to the electron EDM 
is the $H_i^0$-$\gamma$-$\gamma$ Barr-Zee diagram, where Wino loop gives the main contribution. 
On the other hand, the dominant contribution
to the mercury EDM which originates from the quark EDMs
is $H_i^0$-$g$-$g$ Barr-Zee diagram, and the diagram is independent of Wino contribution.
In Figs.~\ref{fig:anatomy-de} and \ref{fig:anatomy-dcd}, 
we show how each contribution to the electron EDM and to the down quark cEDM is decoupled 
for a larger value of $M_2$.


\subsection{DM-nucleon scattering cross section}

Since we are working in the Higgs funnel scenario, 
the DM candidate couples to neutral scalar bosons.
The DM candidate and nucleon interact with each other through these couplings.
The couplings are rather small because of the funnel scenario.
However, the couplings lead to a significant size of the
spin-independent cross section which is  within future prospects of the DM direct detection experiments.

There is also a $Z$-exchange diagram that generates spin-dependent cross section. 
This coupling depends on the mixing between Bino-Higgsino and Bino-Wino in the Bino-like DM scenario.
The mixings are suppressed by the soft breaking neutralino mass parameters.
We find that this coupling is so small that the resultant spin-dependent cross section $\sigma_{SD} = {\cal O}(10^{-8})$~pb
 is smaller than the prospect~\cite{1606.07001}.
In the following, we focus on the spin-independent cross section.

\begin{figure}[tb]
\includegraphics[width=0.45\hsize]{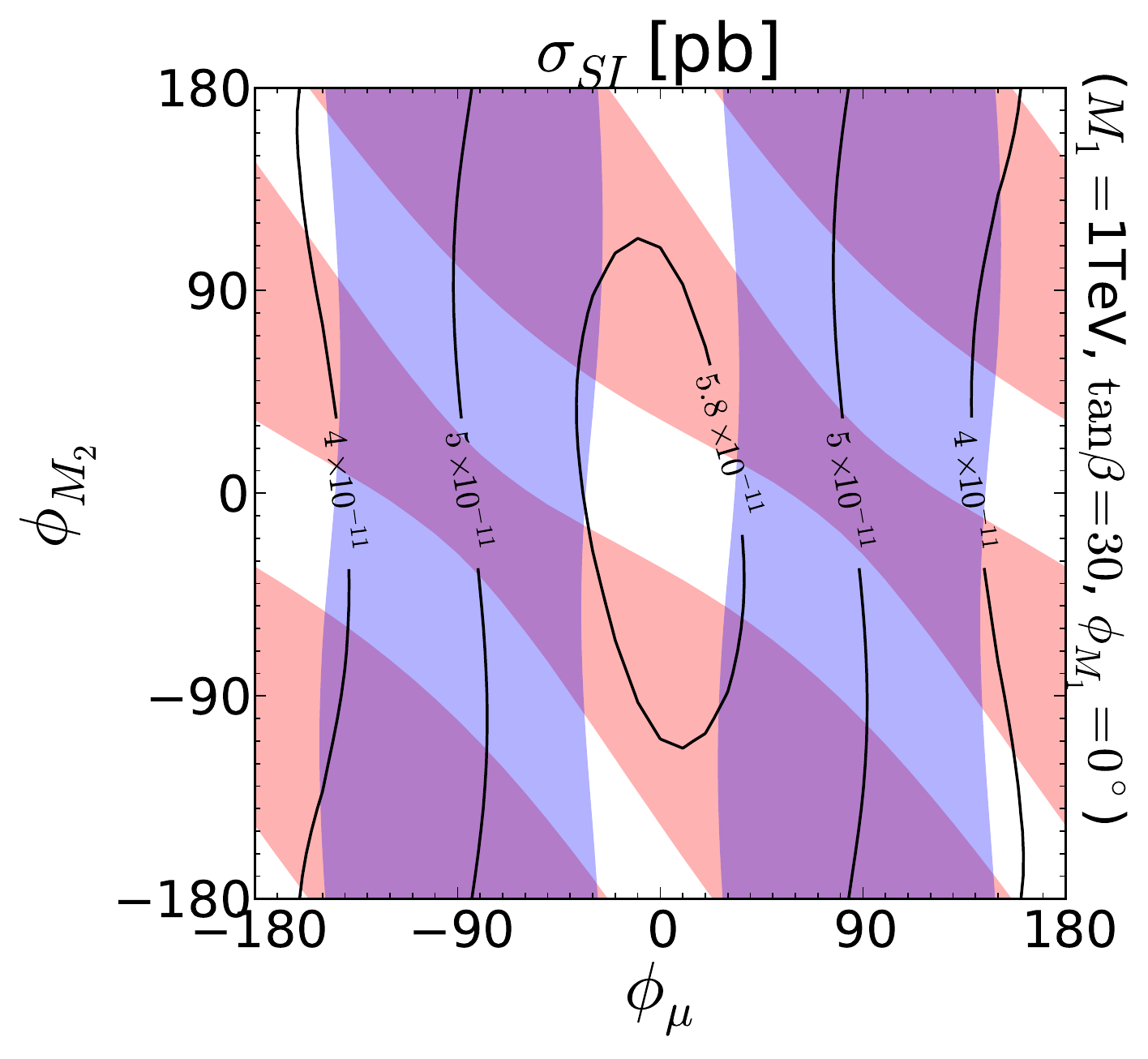}
\includegraphics[width=0.45\hsize]{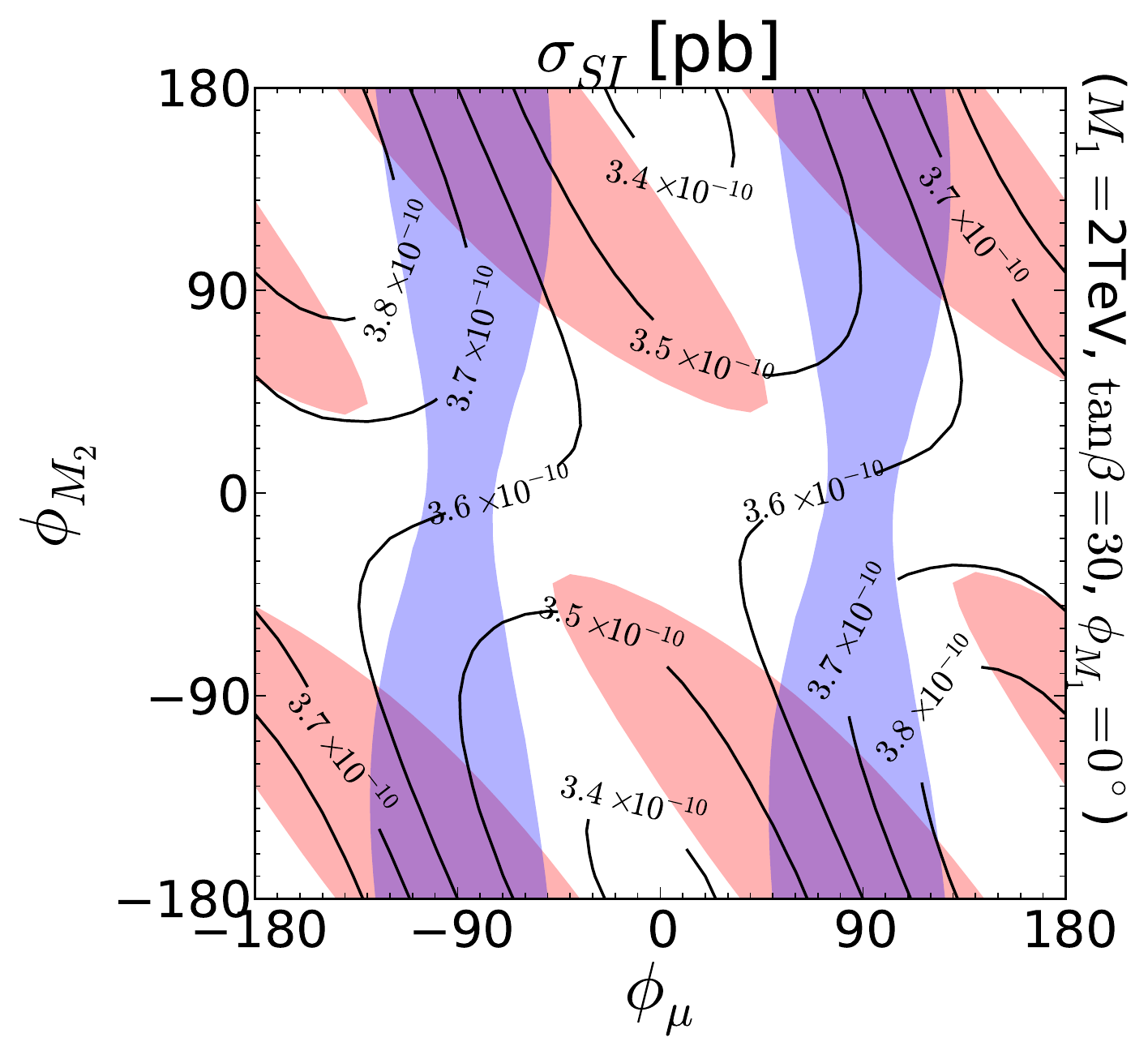}
\caption{
The DM-nucleon scattering cross sections for $\tan\beta=30$ and $\phi_{M_1}=0^\circ$.
The left (right) panel is for $|M_1|=1$~TeV ($2$~TeV). 
The shadings are the same as in Fig.~\ref{fig:edm_mu_vs_M2_with_fiM1-0}.
% Dashed lines show the negative values.
% PandaX constraint on $\Sigma_{\text{SI}}$ is $1.1 \times 10^{-9}$[pb].
% The LZ prospect is $3.3 \times 10^{-11}$[pb].
% prospect[1310.8327]
% Prospect SDp,n by LZ is 0.000011[pb], $6.9 \times 10^{-6}$[pb].
}
\label{fig:xsec1}
\end{figure}

\begin{figure}[tb]
\includegraphics[width=0.45\hsize]{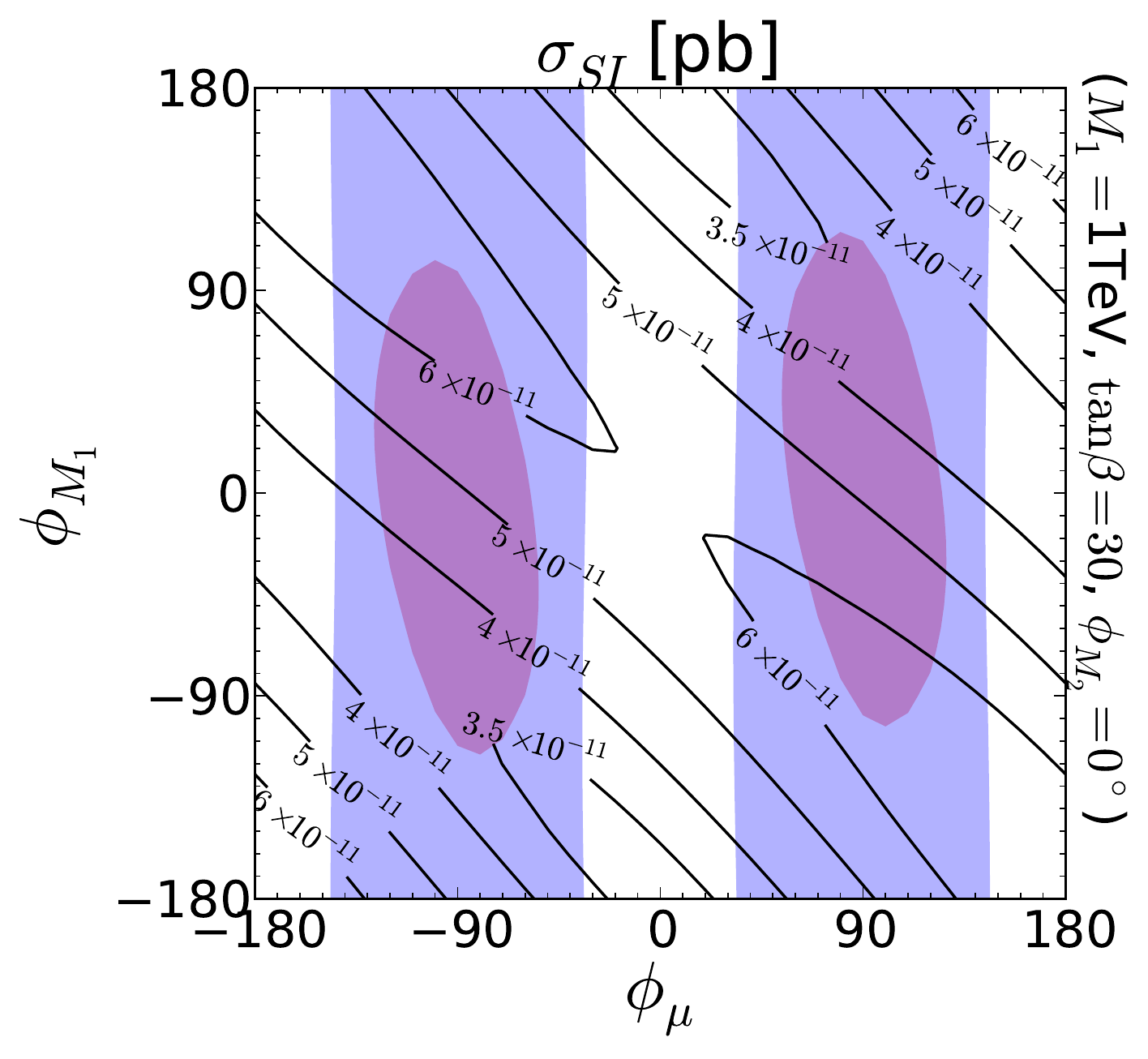}
\includegraphics[width=0.45\hsize]{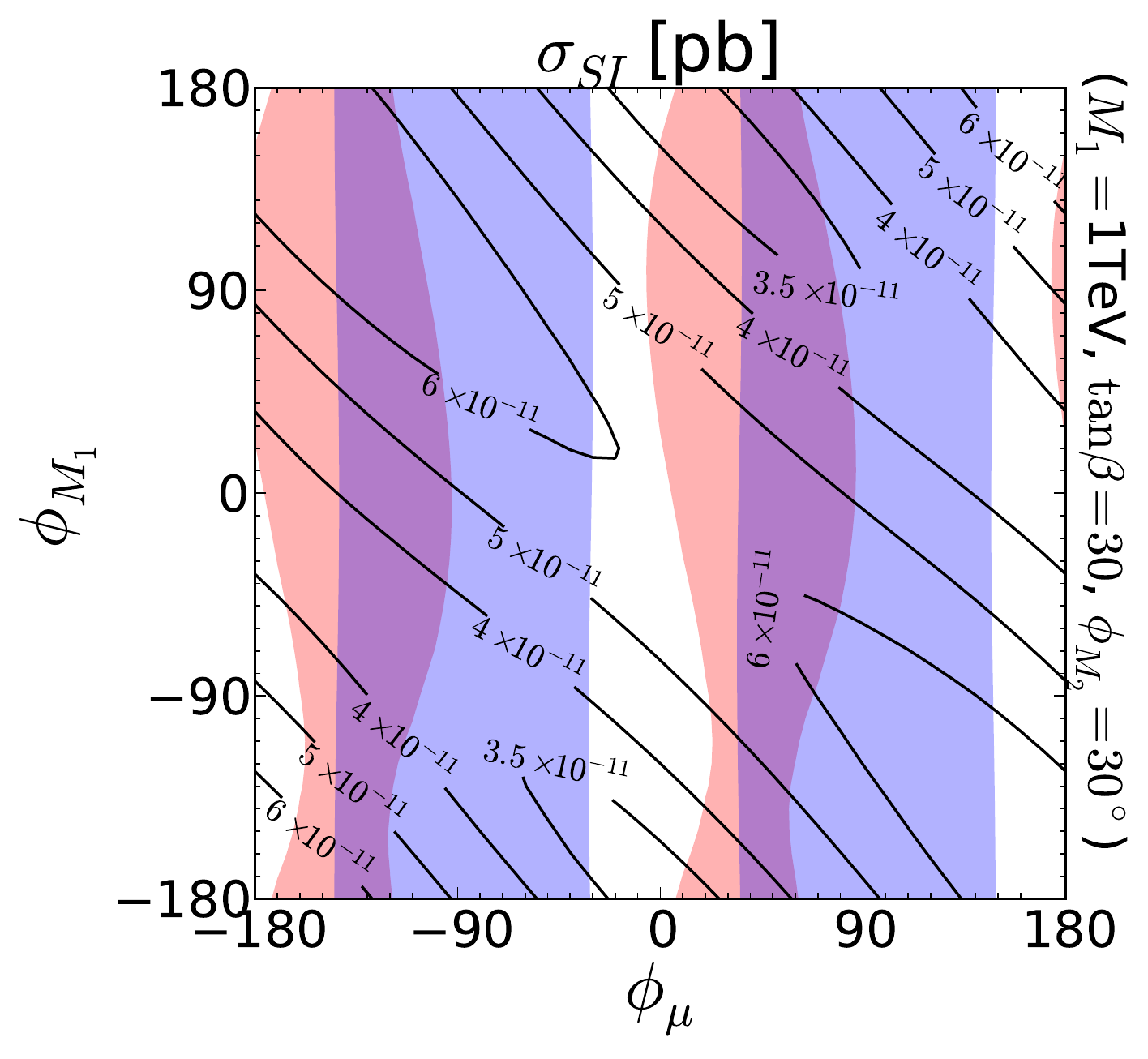}
\caption{
The DM-nucleon scattering cross sections for $M_1=1$~TeV and $\tan\beta=30$.
The left (right) panel is for $|\phi_{M_2}|=0^\circ$ ($30^\circ$). 
The shadings are the same as in Fig.~\ref{fig:edm_mu_vs_M2_with_fiM1-0}.
}
\label{fig:xsec2}
\end{figure}

Figure~\ref{fig:xsec1} shows the $\phi_{M_2}$ and $\phi_{\mu}$ dependence of $\sigma_{\text{SI}}$ where
 its parameter choice is the same as in Fig.~\ref{fig:edm_mu_vs_M2_with_fiM1-0}.
Figure~\ref{fig:xsec2} shows the $\phi_{M_1}$ and $\phi_{\mu}$ dependence of $\sigma_{\text{SI}}$
with the same parameter choice as in Fig.~\ref{fig:M1-dependence}.
We find that the spin-independent cross section is smaller than the current upper 
bound~\cite{Akerib:2016vxi,1705.06655,1708.06917} in all the region of the parameter space
but within the prospects of the DARWIN~\cite{1606.07001},
 the DarkSide-20k~\cite{Aalseth:2017fik}, and the LZ~\cite{Akerib:2018lyp}. 
We also find that 
the scattering cross section 
depends on $\phi_{M_1}+\phi_{\mu}$,
and the $\phi_{M_2}$ dependence is not important.
Since $|M_2|$ is much larger than $|M_1|$ and $|\mu|$ in our analysis,
the sector related to dark matter physics is approximately the Bino-Higgsino system,
and thus the scattering cross section weakly depends on $\phi_{M_2}$.
In the Bino-Higgsino system, there is only one physical CP phase. This is the reason
why the scattering cross section depends on one combination of the CP phases, $\phi_{M_1}+\phi_{\mu}$.

\begin{table}
\caption{Prospects of sensitivity of the spin-independent cross section measurements in	 future experiments.\label{tab:spinindependent}}
\begin{tabular}{c|c|c|c}
$m_{\text{DM}}$&LZ&DARWIN&DarkSide20k\\ \hline
1000~GeV&$1.9\times 10^{-11}$~pb&$3.0\times 10^{-12}$~pb&$1.2\times 10^{-11}$~pb\\  \hline 
2000~GeV&$3.7\times 10^{-11}$~pb&$5.3\times 10^{-12}$~pb&$2.3\times 10^{-11}$~pb \\ \hline 
\end{tabular}
\end{table}
In Table~\ref{tab:spinindependent}, the future prospects of the spin-independent cross section 
measurements are shown.
One finds that all the parameter regions in Figs.~\ref{fig:xsec1} and \ref{fig:xsec2} 
are within the sensitivity of these experiments.

\section{Summary}
\label{SecSummary}

In this paper, we have estimated EDMs of the electron, the neutron, and the mercury as well as 
the DM-nucleon scattering cross section and
 shown the present constraints and prospects 
 in the Bino-like neutralino DM with the heavy Higgs funnel scenario in the CP-violating MSSM.
In our analysis, we have fixed soft SUSY breaking parameters of stops
 to be $\mathcal{O}(10)$~TeV and $\tan\beta=30$ to reproduce the measured SM-like Higgs boson mass and
 other sfermion masses to be $100$~TeV in order to be decoupled from low energy observables.

With such SUSY particle mass spectrum,
 we have shown that CP violating phases of $\mathcal{O}(10)^\circ$
 in the gaugino and Higgsino mass parameters are currently allowed.
Future experiments will be able to constrain those phases at $\mathcal{O}(1)^\circ$ level if the results are null.
We also pointed out that those EDMs have different phase dependence.
For instance, the electron EDM mostly depends on one combination $\phi_{M_2}+\phi_{\mu}$,
 while the neutron and mercury EDMs do on mostly $\phi_{\mu}$ and weakly $\phi_{M_2}$.
Once a few non-vanishing EDMs will be measured,
 it is possible to estimate individual CP phases.
Let us comment on the CP phase of $A_t$. 
In our benchmark points with the similar size of soft stop masses and the trilinear parameter,
 two stop masses are relatively close
 so that the contribution of $\phi_{A_t}$ is suppressed enough for satisfying the 
current experimental bound. 
It should be noticed that even such a suppressed contribution can be tested at the future experiments. 
In the case of a large stop mass spliting,
 which is often accepted in the literature to realize the Higgs boson mass with lighter light stop mass than that we considered here, 
the contribution can be more significant.
In such a case, we will need another observables to determine the individual CP phases.

We also have calculated the dependence of spin-independent
 cross section of the DM in our scenario.
In fact, the non-vanishing CP violation effects change the cross section just by a factor. 
The predicted scattering cross section with a nucleon is
 within the sensitivity of future experiments. 
 
Let us consider the future prospect of our scenario. 
We may expect a positive signal in the direct detection of the DM, which 
provides us an information of the DM mass. 
In our scenario, the extra Higgs bosons should be twice as heavy as the DM, 
so that the heavy Higgs search at LHC can test the scenario.
If the DM mass, cross section, and the heavy Higgs masses are consistent with 
our scenario, we can explore the detail of SUSY breaking sector by EDM experiments
even if the SUSY particles besides the DM are too heavy to be directly discovered 
at the future collider experiments.

\section*{Acknowledgments}
This work was supported by JSPS KAKENHI Grant Numbers 16K17715~[T.A.] and 17H05408~[T.S.].
This work of T.S. was also supported in part by Kogakuin University Grant for the project research.
The work of N.O was supported in part by JSPS Grant-in-Aid for JSPS Fellows, No. 18J10908.

\end{document}